\documentclass[12pt]{article}
\usepackage{graphicx}

\def\beq{\begin{equation}}
\def\eq{\end{equation}}
\def\eeq{\end{equation}}
\def\bea{\begin{eqnarray}}
\def\eea{\end{eqnarray}}

\newcommand{\gsim}{\lower.7ex\hbox{$\;\stackrel{\textstyle>}{\sim}\;$}}
\newcommand{\lsim}{\lower.7ex\hbox{$\;\stackrel{\textstyle<}{\sim}\;$}}

\newcommand{\mc}{\mathcal}
\newcommand{\er}[1]{(\ref{eq:#1})}

\def\centeron#1#2{{\setbox0=\hbox{#1}\setbox1=\hbox{#2}\ifdim
\wd1>\wd0\kern.5\wd1\kern-.5\wd0\fi
\copy0\kern-.5\wd0\kern-.5\wd1\copy1\ifdim\wd0>\wd1
\kern.5\wd0\kern-.5\wd1\fi}}
\def\ltap{\;\centeron{\raise.35ex\hbox{$<$}}{\lower.65ex\hbox{$\sim$}}\;}
\def\gtap{\;\centeron{\raise.35ex\hbox{$>$}}{\lower.65ex\hbox{$\sim$}}\;}
\def\gsim{\mathrel{\gtap}}
\def\lsim{\mathrel{\ltap}}

\def\l{\ell}
\def\slepton{\tilde{\l}}
\def\chii0{\chi_i^0}
\def\chij0{\chi_j^0}
\def\chiipm{\chi_i^{\pm}}

\def\beq{\begin{equation}}
\def\eq{\end{equation}}
\def\eeq{\end{equation}}
\def\barray{\begin{eqnarray}}
\def\earray{\end{eqnarray}}

\def\too{\hookrightarrow}

\def\stau{\widetilde{\tau}}
\def\squark{\widetilde{q}}
\def\sneutrino{\widetilde{\nu}}
\def\sb{\tilde{b}}

\def\st{\tilde{t}}

\def\l{\ell}

\begin{document}

\begin{titlepage}

\begin{center}
\vspace*{-1cm}

\hfill LAUR-08-07771 \\
\hfill RU-NHETC-2008-26 \\
\vskip 0.75in
{\LARGE \bf Probing Supersymmetry With }\\
\vskip .1in
\vspace{.1in}
{\LARGE \bf Third-Generation Cascade Decays } \\
\vskip .1in

\vskip 0.3in
{\large  Michael Graesser}$^1$ {\rm and}  {\large Jessie Shelton}$^2$

\vskip 0.2in

$^1${\em Theory Division, T-2 \\
Los Alamos National Laboratory \\
Los Alamos, NM 87545}

\vskip 0.1in

$^2${\em 
Department of Physics \\
Rutgers University \\
Piscataway, NJ 08854}

\vskip 0.4in

\end{center}

\baselineskip=16pt

\begin{abstract}

\noindent

The chiral structure of supersymmetric particle couplings involving
third generation Standard Model fermions depends on left-right squark
and slepton mixings as well as gaugino-higgsino mixings.  The shapes and 
intercorrelations of invariant mass distributions of a first or second 
generation lepton with bottoms and taus arising from adjacent 
branches of SUSY cascade decays are shown to be a sensitive probe of 
this chiral structure.  All possible cascade decays
that can give rise to such correlations within the MSSM are considered.  
For bottom-lepton correlations the distinctive structure of the
invariant mass distributions distinguishes between decays originating
from stop or sbottom squarks through either an intermediate chargino
or neutralino.  For decay through a chargino the spins of
the stop and chargino are established by the form of the
distribution.  When the bottom charge is signed through soft muon tagging, 
the structure of the same-sign and opposite-sign invariant mass
distributions depends on a set function of left-right and
gaugino-higgsino mixings, as well as establishes the spins of all the
superpartners in the sequential two-body cascade decay.  
Tau-lepton and tau-tau invariant mass distributions arising from MSSM
cascade decays are likewise systematically considered with particular
attention to their dependence on tau polarization.  
All possible
tau-lepton and tau-tau distributions are plotted using a semi-analytic
model for hadronic one-prong taus.
Algorithms for fitting tau-tau and tau-lepton distributions to data are 
suggested.

\end{abstract}

\end{titlepage}

\baselineskip=17pt

\newpage


\section{Introduction}

As the LHC prepares to uncover the physics responsible for electroweak 
symmetry breaking, third-generation fermions present especially interesting 
possibilities as their large
Yukawa couplings make them uniquely sensitive to details of chiral
physics at the electroweak scale.
Moreover, in the context of SUSY, there are several reasons to
anticipate that SUSY signals may be substantially third-generation
enriched.  The LEP2 bounds on the masses of MSSM Higgs bosons
\cite{LEP2} suggest that $\cos 2 \beta \simeq 1$, to maximize the
tree-level contribution to the Higgs mass, favoring moderate or larger
$\tan \beta$.  At large $\tan \beta$, the enhanced Yukawa couplings of
the $b $ and $\tau $ contribute negatively to the running of the
sbottom and stau soft masses, so that sbottoms, staus, and stops all
tend to be lighter than the other sfermions. Large Yukawas also enhance the
coupling of sfermions to higgsinos, thereby increasing the sensitivity
of sparticle decays to gaugino-higgsino mixing.  In addition, large $\tan\beta$ can
lead to large left-right sbottom and stau mixing, further lowering 
sbottom and stau masses
\cite{Baeretal}.  Light stops and large sfermion mixing
are also motivated by the desire to minimize the tuning in the quantum
corrections to the Higgs mass \cite{Perelstein:2007nx}.
Light third-generation sfermions lead to a
significant enrichment of third-generation final states in the signal
\cite{Baeretal}.

Third-generation fermions have complicated in-detector decays which
require more effort to identify and understand.  On the other hand, these
complicated decays allow for measurement of interesting properties
such as polarization (tops, taus) and charge (bottoms) which are
inaccessible for the lighter fermions.  Exploiting these properties
allows for detailed measurement of various aspects of the Lorentz and
chiral structure of new physics.

In this paper we consider SUSY cascade decays which produce
third-generation fermions in the final state.  Our interest here is to
survey the space of possibilities.  Decades of work on supersymmetry 
breaking and its mediation to the Standard Model have led to a
dizzying variety of predictions for possible sparticle spectra.  At the
same time, the continuing absence of deviations from the Standard
Model predictions gives no hints as to which, if any, of these spectra
may be preferred.  
In this paper we enumerate all possible patterns of
invariant mass distributions for $b $-$\ell$, $\l$-$\tau$, and
$\tau$-$\tau$ pairs which arise from adjacent legs of on-shell SUSY
cascade decays, and demonstrate how the special properties of $b$'s
and $\tau$s can be used to obtain further information about the chiral
properties of the MSSM.  Our analysis is model-independent insofar as
it is independent of the spectrum of superpartner masses; we assume
only the existence of on-shell decay modes, the Lorentz structure of 
the SUSY vertices, and the field
content of the MSSM \footnote{We make in addition a few mild
theoretical assumptions, namely: we neglect the electron and
muon Yukawa couplings; we neglect SUSY flavor-violating processes;
we assume lepton flavor universality holds for electrons and muons.  
It is interesting to explore what happens
to the intercorrelations among invariant mass distributions from
cascade decays when these mild assumptions are relaxed \cite{gst1}, but that
is beyond the scope of the present work.}.  This type of model-independent
analysis has been applied to related cascade decay
signals in \cite{burns}.

Two-step on-shell SUSY cascades can give three different fundamental
shapes for invariant mass distributions \cite{barr}.  Intermediate
scalars, as in the decay chains
\beq
\label{eq:basic-triangle}
\chii0\to\l^\pm\slepton ^\mp\to \l^\pm\l ^\mp\chij0
\eeq
give a triangular distribution for the invariant mass of the two
visible standard model fermions,
\beq
\label{eq:triangle}
\frac{1}{\Gamma} \frac{d\Gamma} {dx} = 2 x .
\eeq
Here we have defined the rescaled variable
\[
x\equiv\frac{m_{ff}}{m_{ff,max}}.
\]
Intermediate fermions, as in the decay chains
\beq
\label{eq:blchain}
 \tilde{b}_{L,R} \to b \chii0 \to b \ell \slepton_{L,R}, b \ell \slepton_{R,L}
\eeq
yield ``humps''
\beq
\label{eq:hump}
\frac{1}{\Gamma} \frac{d\Gamma} {dx} = 4 x (1-x ^ 2) \equiv H (x)
\eeq
and ``half-cusps''
\beq
\label{eq:cusp}
\frac{1}{\Gamma} \frac{d\Gamma} {dx} = 4 x^ 3 \equiv C (x)
\eeq
for processes without and with a helicity flip on the intermediate
fermion propagator, respectively. These are the only possibilities
when the intermediate particle is on-shell and the two Standard Model
fermions to be combined together are adjacent in the decay chain
\cite{barr}.

We categorize the $b $-$\ell$ distributions which can arise from
adjacent legs of SUSY cascade decays.  In the MSSM, $b $-$\ell$ final
states can be produced from either the decay of a stop through a
chargino or a sbottom through a neutralino to sleptons.  The patterns
of $b $-$\ell$ invariant mass distributions arising from these cascade
decays have a distinctive structure which discriminates between stop
and sbottom initial states and establishes the spin and Dirac nature
of the chargino.  Signing muonically-decaying $b $-quarks using the
associated soft muon reveals an additional layer of structure which
serves to establish the spin and Majorana nature of the neutralino, as
well as the relative handedness of the sbottom and the slepton
participating in the decay chain.  The ability to establish spins
using signed (for neutralino decay chains) or unsigned (for chargino
decay chains) $b$-$\ell$ distributions are entirely independent of the
existence of a production asymmetry favoring squarks over
anti-squarks.  The shapes and intercorrelations of the $b$-$\ell$
invariant mass distributions are sensitive to both left-right squark
mixing and gaugino-higgsino mixing.

We perform a similar categorization of the $\l$-$\tau$ and
$\tau$-$\tau$ invariant mass distributions which can be realized in
SUSY two-step cascade decays.  As with $b$-$\l$
distributions, the shapes and intercorrelations of $\l$-$\tau$ and
$\tau$-$\tau$ distributions are sensitive to both left-right squark
mixing and gaugino-higgsino mixing.
Invariant mass distributions involving $\tau$s are complicated,
however, by the missing four-momentum of the neutrino coming from the
decay of the $\tau$.  The observable invariant mass distributions
constructed from the $\tau$'s visible hadronic daughters differ
significantly from the underlying triangle, hump, and half-cusp
distributions.  Ditau distributions can nevertheless be used to measure 
superpartner masses with reasonable precision \cite{BaeretalHinchliffeetal,di-tauCMSNOTE}. 
Using the shape of $\tau$-$\tau$ and $\l$-$\tau$ invariant mass distributions 
to further establish more detailed properties of the superpartners 
requires the $\tau$ visible daughter energy spectrum to be carefully 
taken into account.

As the distributions of visible daughter energy depend sensitively on
the polarization of the parent $\tau$
\cite{Tsai:1971vv,Hagiwara:1989fn,Bullock:1991fd,Bullock:1992yt}, the observable
$\l$-$\tau$ and $\tau$-$\tau$ invariant mass distributions are
likewise dependent on the $\tau$ polarization.  Exploiting the
dependence of the visible hadronic daughters on the parent $\tau$
polarization enables a direct probe of the chiral structure of the 
$\tau$ production vertex, opening interesting possibilities both in Higgs physics 
\cite{Bullock:1991fd,Bullock:1992yt,Roy:1991sf,Bullock:1991my,chargedHiggsCMSNote}
and in the MSSM \cite{Nojiri:1994it,Nojiri:1996fp,Choi:2006mt,Mawatari:2007mr,Godbole:2008it}.
Careful study of $\l$-$\tau$ and $\tau$-$\tau$ invariant mass 
distributions arising from cascade decays thereby opens the exciting prospect of
directly probing the mixings and electroweak quantum numbers of the
superpartners participating in the decay\cite{Choi:2006mt,Mawatari:2007mr}.  We use a
semi-analytic approximation to the visible daughter energy spectra for
the hadronic one-prong decay mode, which provides the most sensitive
polarimeter for invariant mass distributions.  Using these spectra, we
plot the possible theoretical $\l$-$\tau$ and $\tau$-$\tau$
distributions that can arise from SUSY cascade decays, and propose
algorithms for fitting these distributions to experimental data.

The outline of the paper is as follows. We begin in Section
\ref{sec:bl} by considering $b$-$\l$ final states.  Section
\ref{sec:tau-nomixing} discusses the possible ditau and lepton-tau
distributions in the limit of no mixing. In Section
\ref{sec:tau-mixing} we turn on mixing, and discuss fitting algorithms
in subsection \ref{sec:rescaling}.  Our conclusions can be found in
Section \ref{sec:conclusion}, and details of our treatment of $\tau$
decay can be found in the Appendix.

\section{$b$-$\ell$ distributions}
\label{sec:bl}

$b$-quarks are useful tools to study SUSY cascade decays.  The
presence of $b$-tags in an event can help serve to separate signal
from background, while the ability to sign semimuonic $b$'s using the
associated muon goes further to help to shed light on the Lorentz
properties of the supersymmetric particles, as we will demonstrate.
We will study the invariant mass distributions of $b$-$\ell$ pairs
which arise from adjacent steps in a cascade decay.

SUSY decay chains which yield adjacent $b$-quarks and leptons are the
decay of sbottoms through a neutralino or stops through a chargino.
Sbottom decays through a neutralino,
\beq
\label{eq:sb}
\sb \to b \chii0\to b\ell\slepton,
\eeq
yield both ``opposite-sign'' ($b^{\pm 1/3} $-$\ell ^{\mp}$) and
``same-sign'' ($b^{\pm 1/3} $-$\ell ^{\pm}$) final states.  In the
limit of zero squark and neutralino mixing, these
processes (\ref{eq:sb}) contribute either opposite-sign humps and
same-sign cusps if the parent sbottom and the final slepton have the
same handedness, or opposite-sign cusps and same-sign humps if the
parent sbottom and the final slepton have the opposite handedness..  
Stop decays through a chargino,
\beq
\label{eq:st}
\st \to b \chiipm\to b\ell\sneutrino ,
\eeq
contribute to opposite-sign final states only with no contribution in
the same-sign channel.  In the limit of zero squark and neutralino
mixing, the distribution in the opposite-sign channel from the process
\er{st} is a hump.

In general events with the decay chain(s) (\ref{eq:sb}, \ref{eq:st})
will also contain additional leptons coming from the subsequent decay
of the sleptons, so there is some ambiguity in selecting which lepton
to pair with the $b$ jet.  Some possible approaches to minimizing this 
combinatoric confusion have been discussed in previous studies of the 
related decay chain $\squark \to q \chii0\to q\ell\slepton$ and its 
UED counterpart 
\cite{barr, Smillie:2005ar, Athanasiou, burns}.  At present we will
concentrate on characterizing the theoretical distribution for the
correct $b$-$\l $ pairing, leaving the question of combinatorics to
future work.

\begin{table}
\begin{center}
\begin{tabular}{ccc}
\hline \hline
& & \\

&  Hump &  Half-Cusp  \\
& &  \\
 \hline
 & & \\

Opposite-Sign & $\sb_L ^\pm \to b ^\pm \chii0 $
                       & $\sb_R ^\pm \to b ^\pm \chii0$ \\

                  & $~~~~~~~~~~~\too  b ^\pm \l ^\mp \slepton_L ^\pm $
                       & $~~~~~~~~~~~\too  b ^\pm \l ^\mp \slepton_L ^\pm $  \\

& &\\

              & $\st_L ^\pm \to b ^\mp\chiipm $  &\\

              & $~~~~~~~~~~~\too b ^\mp \l ^\pm\sneutrino_L $ &   \\

 & & \\

Same-Sign & $ \sb_R ^\pm \to b ^\pm \chii0 $
                       & $ \sb_L ^\pm \to b ^\pm \chii0 $  \\

                  & $~~~~~~~~~~~\too  b ^\pm \l ^\pm\slepton_L ^\mp $
                       & $~~~~~~~~~~~\too  b ^\pm \l ^\pm\slepton_L ^\mp $   \\

 & &\\

& \\
\hline \hline
\end{tabular}
\caption{Possible invariant mass distributions for the $b$-$\l$
final states.  Here by a
slight abuse of notation $\pm$ distinguishes between (s)quark and
anti-(s)quark, so that $b ^\pm$ has charge $\pm 1/3$.  Distributions
for final state $\slepton_R $ are obtained by exchanging hump and
half-cusp. Squark right-left mixing is neglected and neutralinos and
charginos are taken to be pure gaugino.
\label{table:bl-unmix}}
\end{center}
\end{table}

The possible SUSY $b $-$\l$ distributions in the limit of no squark or
neutralino mixing are summarized in Table~\ref{table:bl-unmix} 
\footnote{In the name of generality, we remark that interchanging the 
role of the squark and slepton does not alter the shapes and 
correlations of the $b$-$\l$ distributions, so that (e.g.) the decay chain
$\sb_L ^\pm \to b ^\pm \chii0 \to b ^\pm \l ^\mp \slepton_L ^\pm $
yields the same $b$-$\l$ invariant mass distribution as the flipped decay chain
$\slepton_L ^\pm \to \l ^\pm \chii0 \to \l ^\pm b ^\mp \sb_L ^\pm $.}.  If
(e.g.) the decay chain $\sb_L ^\pm \to b ^\pm \chii0 \to b ^\pm \l
^\mp \slepton_L ^\pm $ exists in the signal, then (for a standard
Majorana neutralino) so must the decay chain $\sb_L ^\pm \to b ^\pm
\chii0 \to b ^\pm \l ^\pm \slepton_L ^\mp $.  Therefore there must
exist both a hump distribution in the opposite-sign channel and a cusp
distribution in the same-sign channel, with equal normalizations and
endpoints.  Since the hump and the half-cusp sum to a triangle, if the
opposite-sign and same-sign channels cannot be distinguished, then the
information about the spin of the neutralino is lost.  Thus the
ability to sign the $b$-jet allows for determination both of the spin
of the neutralino and of its Majorana nature.

When the sbottom and the slepton have the same handedness then
opposite sign distributions are humps and same-sign distributions are
half-cusps; the situation is reversed when the sbottom and the slepton
have different handedness.  It is possible to obtain spectra which
allow all four possible decay chains ($\sb_{L, R}\to b\l\slepton_{L,
R} $ and $\sb_{L, R}\to b\l\slepton_{R, L} $) to be realized
simultaneously.  This will complicate spin measurements as the sums of
the overlapping distributions coming from these processes will tend to
wash out the spin correlations.  The greater the mass splitting
between right- and left-handed squarks and/or sleptons, the more
distinct the endpoints of the different distributions will be, and the
easier it will be to disentangle the contributions from different
processes to the total $b $-$\l$ distributions.  The possibility of
making a spin measurement in this channel will also depend on the
relative branching fractions into sleptons of different handedness.
Consider the case when right- and left-handed sleptons are nearly
degenerate.  Then the different gauge quantum numbers of the right-
and left-handed sleptons will still yield (e.g.) $\Gamma (\sb_L \to
b\chii0\to b\l\slepton_L )/\Gamma (\sb_L \to b\chii0\to
b\l\slepton_R)\neq 1$.  The total $b$-$\l $ distribution will then
retain some spin information.

Stops, on the other hand, contribute only to the opposite-sign
channel, so comparison of opposite-sign and same-sign distributions
will be an important tool to disentangle possible contributions from
stop and sbottom squarks.  

When the intermediate chargino is pure gaugino, only decay chains 
beginning from initial $\st_L $ can contribute to $b$-$\l $ final states.
However, as the lepton Yukawas are negligible, it is also sensible 
to consider the case where the chargino is nearly pure higgsino but still 
decays to lepton-slepton pairs as a gaugino. In this case, the $\st_R$ can
decay through an up-type higgsino in the decay chain 
$\st_R ^\pm \to b^\mp\chiipm \to b ^\mp \l ^\pm\sneutrino $,
yielding an opposite-sign hump distribution.  The $\st_L$ can decay 
through a down-type higgsino via the decay chain $ \st_L ^\pm \to b ^\mp\chiipm
\to b ^\mp \l ^\pm\sneutrino $, which yields an opposite-sign half-cusp
distribution.  

In stop decay, the $b$-tag alone suffices to establish that the
intermediate fermion is Dirac, rather than Majorana, and the
additional sign information is not necessary: if the intermediate
fermion were Majorana, then summing the distributions for the final
states $\bar b$-$\l ^-$ and $b$-$\l ^-$ would give a triangle, while
for an intermediate Dirac particle, only one of the helicity states
can contribute, leading to a hump distribution (or a half-cusp distribution 
when the chargino is a down-type higgsino) even when both final 
states are summed.

This is an example of a general point: it is possible to observe
nontrivial (non-triangular) distributions in the absence of sign
information, if a symmetry forbids one channel from contributing (as
for stop decays through charginos, here, or as for Dirac neutralinos
\cite{gst1}).  Nontrivial distributions also
can be obtained if the two channels contribute with unequal weights,
as is the case when a production asymmetry favors squarks over
anti-squarks \cite{barr}.  However, event-by-event signing of the
semimuonic $b$'s allows a direct observation of any angular
correlations coming from intermediate Majorana neutralinos,
independent of any possible production asymmetry.  $b$-jet signing as
a tool to improve spin measurements at the LHC has been mentioned in
\cite{Alves:2006df, Wang:2006hk}.

We now go on to discuss how this story is modified in the presence of
nontrivial squark and neutralino mixing.

\subsection{$b$-$\l$ distributions with nontrivial squark and neutralino mixing}

Left-right sfermion mixing renders the SUSY sfermion-fermion-gaugino
vertices less chiral and thereby alters the observable kinematical
distributions \cite{Goto:2004cpa}.
Similarly, higgsino interactions proportional to a fermion Yukawa
coupling involve the opposite chirality of the fermion relative to
gaugino interactions.  Sfermion and neutralino mixings, then, have a
qualitatively similar effect on the invariant mass distributions.  In
this section we will detail the sensitivity of difermion invariant
mass distributions to both sfermion mixing and neutralino mixing,
including both effects simultaneously.  While here we concentrate on
$b$-quarks and $b$-$\l$ distributions, the same physics will be
relevant to lepton-tau and ditau distributions in
section~\ref{sec:tau-mixing}.

Consider the hump and half-cusp distributions which arise from decay
chains with an intermediate fermion.  Once mixing is turned on, both
helicity states of the intermediate fermion can contribute to a given
channel, with relative weights determined by the mixing.  The
observable same-sign and opposite-sign distributions are then a
weighted sum of hump and half-cusp distributions.  Define the squark
left-right mixing angles through
\beq
 \squark_1 =\cos\theta_{\squark} \,\squark_R ^*+\sin\theta_{\squark}\, \squark_L.
\eeq
(As usual, $\squark_1$ is taken to be the lighter of the two squarks.  In
our conventions $q_R$ is a left-handed anti-quark, and thus $\squark_R$
is an anti-squark.)
Define also the unitary matrix $U$ which diagonalizes the neutralino
mass matrix.  The sbottom-bottom-neutralino couplings are then
governed by the interaction Lagrangian
\begin{equation}
{\cal L}= \sb_1 \left( b_R \chii0 \; n^R_{1,i} + (\chii0)^\dag t_L ^\dag \; n^L_{1,i} \right)
         + \sb_2 \left( b_R \chii0 \; n^R_{2,i} + (\chii0)^\dag t_L ^\dag \; n^L_{2,i} \right)
        + \mathrm{H.c.}
\end{equation}
with parameters
\barray
n^R_{1, i} & = & \sin\theta_{\sb}\lambda_b U_{di}^* +\cos\theta_{\sb}{\sqrt 2 g'\over 3} U_{Bi}^*\\
n^L_{1, i} & = & \sin\theta_{\sb}\left( -{g\over\sqrt 2} U_{iW} +{g'\over 3\sqrt 2} U_{iB }\right) +\cos\theta_{\sb} \lambda_b U^*_{di}  \\
n^R_{2, i} & = & \cos\theta_{\sb}\lambda_b U_{di}^*-\sin\theta_{\sb}{\sqrt 2 g'\over 3} U_{Bi}^*, \\
n^L_{2, i} & = & \cos\theta_{\sb}\left( -{g\over\sqrt 2} U_{iW} +{g'\over 3\sqrt 2} U_{iB }\right) -\sin\theta_{\sb} \lambda_b U_{di}^*
\earray
Note that $n^L_{1,i} \to 0$, $n^R_{2, i}\to 0$ as both mixings are
turned off, that is, as $\theta_{\sb}, U_{di} \to 0$, while $n^L_{2,
i}$, $n^R_{1,i}$ remain finite.  The index $i$ specifies the
neutralino mass eigenstate $i$, while the indices $d, B, W$ run over
the gauge eigenstates (here down-type higgsino, bino, and wino,
respectively).  We now define the angles
\beq
\label{eq:b-neutralino-angle}
\cos ^ 2\alpha_{1i}\equiv {| n^R_{1, i} | ^ 2\over | n^R_{1, i} | ^ 2 + | n^L_{1, i} | ^ 2} ~~,~~
\cos ^ 2\alpha_{2i}\equiv {| n^L_{2, i} | ^ 2\over | n^L_{2, i} | ^ 2 + | n^R_{2, i} | ^ 2}\; ;
\eeq
in the limit of zero mixing, $\cos\alpha_{1i}$ and $\cos\alpha_{2i}$
both go to unity.  Decays of sbottoms through the neutralino
$\tilde{\chi}^0_i$ will be weighted by these angles.  Meanwhile, the
stop-bottom-chargino couplings are governed by the interaction
Lagrangian
\begin{equation}
{\cal L}= \st_1 \left( b_R\chi_i^- \; c^R_{1, i} + (\chi_i ^ +) ^\dag b_L ^\dag \; c^L_{1, i} \right)
         + \st_2 \left( b_R\chi_i^- \; c^R_{2, i} + (\chi_i ^ +) ^\dag b_L ^\dag \; c^L_{2, i} \right)
 + \mathrm{H.c.}
\end{equation}
with parameters
\barray
c^L_{1, i} & =  & -\sin\theta_{\st} g V_{iW} + \cos\theta_{\st}\lambda_t V_{iu} \\
c^R_{1, i} & = &\sin\theta_{\st}\lambda_b W^*_{di}\\
c^L_{2, i} & =  & -\cos\theta_{\st}  g V_{iW}- \sin\theta_{\st}\lambda_t V_{iu} \\
c^R_{2, i} & =  & \cos\theta_{\st}\lambda_b W^*_{di},
\earray
The unitary matrix $V$ (not the CKM matrix!) diagonalizes the
positively-charged left-handed charginos $(\tilde{W}^+,
\tilde{h}^+_u)$, and the unitary matrix $W$ diagonalizes the
negatively-charged left-handed charginos $(\tilde{W}^-,\tilde{h}_d^-)$.
We then define the angles
\beq
\label{eq:b-chargino-angle}
\cos ^ 2\beta_{1 i}\equiv  {| c^R_{1, i} | ^ 2\over | c^R_{1, i} | ^ 2 + | c^L_{1, i} | ^ 2} ~~,~~
\cos ^ 2\beta_{2 i}\equiv  {| c^L_{2, i} | ^ 2\over | c^L_{2, i} | ^ 2 + | c^R_{2, i} | ^ 2} \; ,
\eeq
where again $\cos\beta_{1i}$ and $\cos\beta_{2i}$ go to unity in the
limit of zero mixing.  Decays of stops through the chargino
$\tilde{\chi}^+_i$ will be weighted by these angles.
Table~\ref{table:bl} summarizes the $b $-lepton distributions in the
presence of mixing; a similar approach to mixing was taken in
\cite{burns}.

\begin{table}
\begin{center}
\begin{tabular}{ccccc}
\hline \hline
& & & & \\

&  Hump & & Half-Cusp & \\
& Process & Weight & Process & Weight \\
& &  & & \\
 \hline
 & & & \\

              & $ \sb_1 ^\pm \to b ^\pm \chii0 $ &
                       & $\sb_1 ^\pm \to b ^\pm \chii0 $ & \\

                  & $~~~~~~~~~~~\too b ^\pm \l ^\mp \slepton_L ^\pm $  & $ \sin ^ 2\alpha_{1 i} $
                       & $~~~~~~~~~~~\too b ^\pm \l ^\mp \slepton_L ^\pm $  & $\cos ^ 2\alpha_{1 i}$ \\
                  & $~~~~~~~~~~~\too b ^\pm \l ^\mp \slepton_R ^\pm $  & $ \cos ^ 2\alpha_{1 i} $
                       & $~~~~~~~~~~~\too b ^\pm \l ^\mp \slepton_R ^\pm $  & $\sin ^ 2\alpha_{1 i}$\\

Opposite-Sign & $\sb_2 ^\pm \to b ^\pm \chii0 $ &
                       & $\sb_2 ^\pm \to b ^\pm \chii0$ & \\

                  & $~~~~~~~~~~~\too  b ^\pm \l ^\mp \slepton_L ^\pm $  & $ \cos ^ 2\alpha_{2 i} $
                       & $~~~~~~~~~~~\too  b ^\pm \l ^\mp \slepton_L ^\pm $  & $ \sin ^ 2\alpha_{2 i}$ \\
                  & $~~~~~~~~~~~\too b ^\pm \l ^\mp \slepton_R ^\pm $  & $ \sin ^ 2\alpha_{2 i}$
                       & $~~~~~~~~~~~\too b ^\pm \l ^\mp \slepton_R ^\pm $  & $ \cos ^ 2\alpha_{2 i} $ \\

              & $\st_{1,2} ^\pm \to b ^\mp\chiipm $ &
                       & $ \st_{1,2} ^\pm \to b ^\mp \chiipm $  &\\

              & $~~~~~~~~~~~\too b ^\mp \l ^\pm\sneutrino_L $ & $ \cos ^ 2\beta_{(1,2) i} $
                       &  $~~~~~~~~~~~\too  b ^\mp \l ^\pm\sneutrino_L$ & $ \sin ^ 2\beta_{(1,2) i}$  \\

 & & & &\\
\hline
 & & & &\\

          & $ \sb_1 ^\pm \to b ^\pm \chii0 $ &
                       & $ \sb_1 ^\pm \to b ^\pm \chii0 $ & \\

                  & $~~~~~~~~~~~\too b ^\pm \l ^\pm\slepton_L ^\mp $  & $\cos ^ 2\alpha_{1 i}$
                       & $~~~~~~~~~~~\too b ^\pm \l ^\pm\slepton_L ^\mp $  & $\sin ^ 2\alpha_{1 i}$ \\
                  & $~~~~~~~~~~~\too b ^\pm \l ^\pm\slepton_R ^\mp $  & $ \sin ^ 2\alpha_{1 i}$
                       & $~~~~~~~~~~~\too b ^\pm \l ^\pm\slepton_R ^\mp $  & $\cos ^ 2\alpha_{1 i}$\\

Same-Sign & $ \sb_2 ^\pm \to b ^\pm \chii0 $ &
                       & $ \sb_2 ^\pm \to b ^\pm \chii0 $ & \\

                  & $~~~~~~~~~~~\too  b ^\pm \l ^\pm\slepton_L ^\mp $  & $\sin ^ 2\alpha_{2 i} $
                       & $~~~~~~~~~~~\too  b ^\pm \l ^\pm\slepton_L ^\mp $  & $ \cos ^ 2\alpha_{2 i}$ \\
                  & $~~~~~~~~~~~\too b ^\pm \l ^\pm\slepton_R ^\mp $  & $ \cos ^ 2\alpha_{2 i} $
                       & $~~~~~~~~~~~\too b ^\pm \l ^\pm\slepton_R ^\mp $  & $\sin ^ 2\alpha_{2 i}$\\

 & & & &\\

& \\
\hline \hline
\end{tabular}
\caption{Distributions for the $b$-$l$ invariant mass including both
gaugino-higgsino mixing and third-generation squark mixing. Here by a
slight abuse of notation $\pm$ distinguishes between (s)quark and
anti-(s)quark, so that $b ^\pm$ has charge $\pm 1/3$. The relative
weights are normalized such that the coefficients for the hump and
cusp distributions in a given channel sum to unity.  The angles
$\alpha_{ki},\beta_{ki} $ are defined in
equations~(\ref{eq:b-neutralino-angle})
and~(\ref{eq:b-chargino-angle}). The endpoints of the distributions will
depend on the masses of the superpartners participating in the cascade.
\label{table:bl} }
\end{center}
\end{table}

Notice that for stop squark decays through charginos, admixture of the
cusp distribution depends on a sizable down-type higgsino component of the
intermediate chargino.  This can be seen as follows.  As the lepton
Yukawas are negligible, the chargino decay to lepton-sneutrino pairs will 
occur through its gaugino component.  To obtain a half-cusp distribution,
the $b$ must then be right-handed (as its charge is fixed).  But as
the wino does not couple to the right-handed $b$, a half-cusp
distribution can only be obtained if the intermediate chargino has a
down-type higgsino component, independent of any possible mixing of the stop
squarks.  The coupling of the higgsino to the $b_R$ depends on
$\lambda_b$ and is therefore enhanced at large $\tan \beta$.  In
short, if the chargino is taken to have no down-type higgsino
component, then stop mixing alone will not alter the observed hump
distribution.  In this context it is useful to recall that no sign
information of the $b$-quark is necessary to observe the hump
distribution coming from stop decay.

Fitting the observed $b$-$\ell$ distributions to sums of humps and
half-cusps thereby measures the mixing parameters of
equations~(\ref{eq:b-neutralino-angle})
and~(\ref{eq:b-chargino-angle}).  These mixing parameters are the only
linear combinations of Lagrangian parameters which can be measured using 
these cascade decays alone.  For stop squark decays, the mixing parameters 
and any production asymmetry can be separately measured by considering the
separate distributions $b$-$\l^+$ and $b$-$\l^-$.  For sbottom squark decays,
the mixing parameters and the production asymmetry can be separately measured 
if the $b$-quark is signed and the distributions for all four possible combinations
of lepton and $b$ signs are independently examined.

\subsection{$b$-jet signing and measurement of mixings}

In practice signing the $b$-quark cannot be done with arbitrary
purity, and there will be a non-trivial but well-characterized
mis-sign rate.  In particular there is an irreducible contribution to
the mis-sign rate coming from oscillations of the parent $b$-quark
inside neutral $B$-mesons, which is of order $12\% $ \cite{pdg}.
Recent simulations indicate that a mis-sign rate of $15.4\% $ can
be achieved, with an efficiency of $1.2\% $ \cite{stevenyi}.  To
measure spin and mixings, $b$-$\ell$ distributions should be fit to
sums of humps and half-cusps.  Let the mis-sign fraction be $F $, and
consider decay of $b $-squarks through neutralinos. If the theoretical
distribution in a given channel is
\[
D_{theory}(x) = f H(x) +(1-f) C(x),
\]
where $f$ and $(1-f)$ are the (sines and cosines of) angles listed in
Table \ref{table:bl}, then the experimentally observed distribution is
\barray
\nonumber
D_{obs}(x) &=& (1-F) \left( f H(x) + (1-f) C(x) \right) + F \left( (1-f) H (x) +f C(x)\right) \\
           &=& \hat f H(x) +(1-\hat f) C (x),
\earray
where
\beq
\label{eq:fhat}
\hat f\equiv f+F-2 fF.
\eeq
Fits to data then directly measure $\hat f$, which through \er{fhat}
measures the mixing parameters of
equation~(\ref{eq:b-neutralino-angle}).  In a realistic situation
there will be a trade-off between purity, that is, minimizing $F $,
and acceptance.  In a full analysis any $p_T $-dependence of $F $ can
be included.

\begin{figure}
\begin{center}
\includegraphics[width=10.0cm]{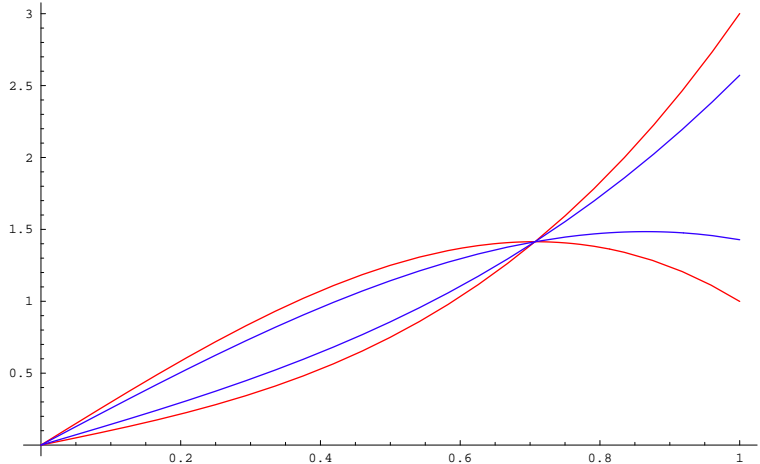}
\end{center}
\vspace{-0.75cm}
\begin{small}
\caption{Observable $b$-$\ell$ invariant mass distributions arising
from $\sb$ decay through a neutralino, for $\cos\theta_b = 0.775$ and
intermediate pure bino.  The outer (red) pair of curves show the
observable distributions assuming a $15\% $ mis-sign rate.  The inner
(blue) pair of curves show the same distributions assuming a $30\% $
mis-sign rate.
\label{fig:bmix}}
\end{small}
\end{figure}

In Figure~\ref{fig:bmix} we plot the observable $b$-$\ell $
distributions for an intermediate pure bino and $\cos\theta_b =
0.775$, assuming no production asymmetry.  Two sets of curves are
shown.  The outer (red) pair are the opposite- and same-sign
distributions which would be observed with a $15\% $ mis-sign rate.
The inner (blue) pair are the opposite- and same-sign distributions
which would be observed with a $30\% $ mis-sign rate.  Even with a
$30\% $ mis-sign rate the deviation of the distributions from each
other and from the triangular distribution is clear, indicating the
presence of an intermediate Majorana fermion.  Decreasing the mis-sign
rate further to $15\% $ significantly enhances the difference between
the two channels, and hence the sensitivity to mixing.

Let us finally mention that in the semi-muonic decay modes of the $b$,
the missing energy carried away by the neutrino alters the energy
distribution of the observed $b $-jets, and a full analysis must
account for this effect.  As the polarization of the $b$-quark is
almost entirely randomized in hadronization \cite{Falk:1993rf}, the
visible spectrum of $b$-quark decay products is not dependent on the
details of the vertex where the $b $ originated. The net effect of the
loss of the neutrino four-momentum can be described by a convolution
of the invariant mass distributions involving the parent $b$-quark
with a calculable universal transfer function which accounts for the
distribution of visible energy in the $b$-jets after the loss of the
neutrino.

\section{Ditau and lepton-tau distributions: no mixings}
\label{sec:tau-nomixing}

As taus decay within the detector via $\tau\to\nu X $, the full
four-momentum of the $\tau $ is not observable, and events involving
final state $\tau $s require careful attention to characterize and
understand.  However, the hadronic decay modes of the $\tau$ do allow
right- and left-handed $\tau $s to be distinguished statistically.  A
left-handed $\tau $ preferentially emits the neutrino parallel to its
direction of motion, resulting in a softer spectrum of visible decay
products, while a right-handed $\tau $ preferentially emits the
neutrino anti-parallel to its direction of motion, resulting in a
harder spectrum of visible decay products.  This difference in the
energy distributions of the $\tau$ daughter products therefore can be
used as a handle on the possible chiral couplings of new physics to
$\tau$ leptons \cite{Hagiwara:1989fn,Bullock:1991fd,Bullock:1992yt}.

When the $\tau$ is highly boosted in the lab frame, as it is in most
cascade decays, its visible decay products $d$ are collinear, and to
good approximation we can take
\[
p_d = z p_\tau,
\]
where $z$ is the fraction of the lab frame $\tau$ energy carried by
the daughters $d$,
\begin{equation}
z = \frac{E_d}{E_{\tau}} ~.
\end{equation}
The quantity $z$ is invariant under boosts along the direction of the
tau.  Also, when the tau is highly boosted, helicity and chirality may
be used interchangeably.  The distribution of $z$ is correlated with
handedness of the $\tau$, as noted above.

At the level of the parent $\tau$s, two-step on-shell SUSY cascades
can lead to triangle, hump, and half-cusp invariant mass distributions.
Intermediate staus, as in the decay chains
\beq
\label{eq:tau-triangle}
\chii0\to\tau_{L, R} ^\pm\stau_{L, R} ^\mp\to \tau_{L, R} ^\pm\tau_{L, R} ^\mp\chij0
\eeq
give triangles, while the decay chains
\beq
\label{eq:lt-chain}
 \slepton_{L,R} \to \ell \chii0 \to\ell\tau \stau_{L, R}, \ell\tau \stau_{R, L}
\eeq
and
\beq
\label{eq:tt-chain}
 \stau_{L,R} \to \tau \chii0 \to\tau\tau \stau_{R,L}
\eeq
yield humps and half-cusps.  For final state $\tau$'s, these
distributions are not directly observable.  Rather, we must convolve
these distributions with the probability $P_{(d)}^\pm (z)$ that a
parent $\tau$ decays to visible daughter particle(s) $d$ with momentum
fraction $z = p_d / p_\tau$.  The probability $P^{\pm}_{(d)}$ depends
on the helicity of the parent $\tau$, denoted $\pm$.  CP invariance
ensures that the energy distributions of a $\tau$ with a given
helicity and its anti-particle are identical. Thus the
negative-helicity $\tau^-$ and the positive-helicity $\tau^+$ have
identical energy distributions, which we denote by $P^-_{(d)}(z)$
throughout.  Similarly, the energy distributions for the
positive-helicity $\tau^-$ and its antiparticle, the negative-helicity
$\tau^+$, are given by $ P^+_{(d)}(z)$.

\begin{figure}
\begin{center}
\includegraphics[width=10.0cm]{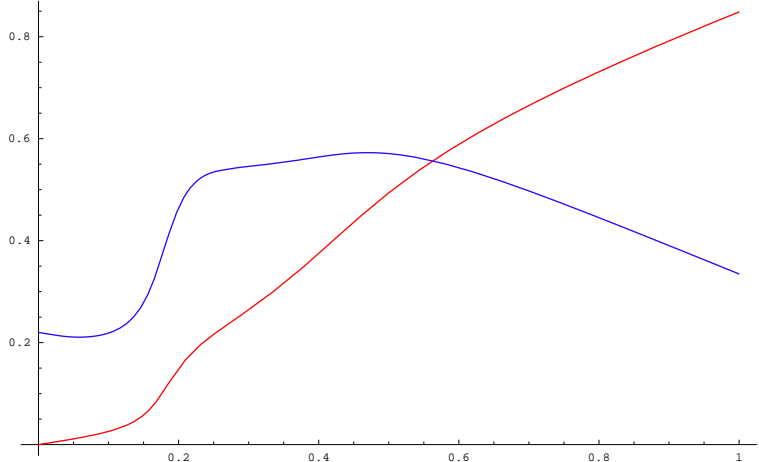}
\end{center}
\vspace{-0.75cm}
\begin{small}
\caption{The $\tau $ daughter energy fraction distributions
$ P^\pm_{(1)}(z)$ for
hadronic one-prong taus, for positive (red) and negative (blue)
helicity taus.
\label{fig:p(z)}}
\end{small}
\end{figure}

Our analysis uses hadronic one-prong taus as they yield the greatest
sensitivity to the tau polarization; our treatment of the one-prong 
decay mode closely follows that of \cite{Bullock:1992yt}.  We model the energy fraction
distributions $ P^\pm_{(1)}(z)$ for one-prong taus by summing the
three dominant contributions to this mode, the decays $\tau^- \to \nu
\pi^ 0$, $\tau^- \to \nu \rho ^-\to\nu \pi^- \pi^ 0$, and $\tau^- \to
\nu a_1 ^-\to \nu \pi^- \pi^ 0 \pi ^ 0$.  Details of this computation
are presented in the Appendix.  The total daughter energy fraction
distributions $P_{(1)}^\pm (z)$ that result are shown in
Figure~\ref{fig:p(z)}.

The functions $P_{(3)}^\pm (z)$ for three-prong taus are comparatively
insensitive to tau polarization.  The three-prong decay mode is
principally due to the decay $\tau \to a_1 \nu $, and the total
visible energy fraction for the three-prong decay mode is
approximately equal to the visible energy fraction contributed by the
$a_1$'s to the one-prong decay mode.  The mass difference between the
$a_1$ and the $\tau $ is not large, and the contributions of
longitudinally and transversely polarized $a_1$ mesons add to a nearly
spin-independent quantity (further details can be found in the
Appendix).  Therefore including the three-prong decay mode does not
increase sensitivity to $\tau$ polarization in invariant mass
distributions\footnote{The three-prong decay modes can, however, be used as
an effective polarimeter by examining the distribution of energy among
the daughter pions, which is sensitive to the polarization of the
$a_1$, and therefore to the helicity of the $\tau $
\cite{Bullock:1991my}.  Cuts on the relative energy distributions of
the daughter hadrons as a way to distinguish between tau polarizations
in SUSY cascades have been discussed in \cite{Godbole:2008it}.}.

In lepton-tau final states, the observable invariant mass variable is
$m_{\ell d}$, where $d$ again denotes the visible $\tau $ decay
product(s).  The variable $m^ 2_{\ell d}$ is distributed according to
\beq
\label{eq:lt-dist}
\frac{1}{\Gamma} \frac{d\Gamma} {d m ^ 2_{\ell d}} =
    \frac{1}{\Gamma} \int_{m ^ 2_{\ell d}}^ 1 \frac{dz}{z} P_{(d)} ^\pm (z)  \left.
    \frac{d\Gamma}{d m ^ 2_{\ell \tau}}\right|_{ m ^ 2_{\ell \tau}=m ^ 2_{\ell d}/z}.
\eeq
In ditau final states, we have similarly
\beq
\label{eq:tt-dist}
\frac{1}{\Gamma} \frac{d\Gamma} {d m ^ 2_{dd'}} =  \frac{1}{\Gamma} \int_{m ^ 2_{d d'}}^ 1
    \int_{m ^ 2_{d d'}/z_1} ^ 1 \frac{dz_1}{z_1} P_{(d)} ^\pm (z_1) \frac{dz_2}{z_2} P_{(d')} ^\pm (z_2)\left.
     \frac{d\Gamma} {d m ^ 2_{\tau \tau}}\right|_{ m ^ 2_{\tau \tau}= m ^ 2_{dd'}/(z_1 z_2)} .
\eeq
Equations (\ref{eq:lt-dist}) and (\ref{eq:tt-dist}) lead to calculable
predictions for the invariant mass distributions of detected tau decay
products which will in general depend on the underlying SUSY process,
the polarization of the parent tau, and the decay mode(s) selected.
We plot invariant mass distributions of this form normalized to unity,
rather than to the $45\%$ branching ratio (or the branching ratio squared)
into the hadronic one-prong decay modes we model.

Our interest in the rest of this paper will be to explore how
information about $\tau$ polarization can be used in conjunction with
invariant mass distributions to further measure properties of a
general SUSY model.  Our principal aim here is to establish the range of 
theoretical possibilities.

For the purposes of the remainder of this section, we assume purely
chiral couplings, that is, we work in the limit of vanishing
left-right stau mixing and Yukawa couplings, and assume that the
neutralinos participating in the cascade decays couple purely as
gauginos.  This is an idealization, and in many well-motivated
scenarios left-right stau mixing is appreciable. First we will
consider ditau final states, then in section \ref{sec:lepton-tau} move
onto lepton-tau final states. Then in section \ref{sec:tau-mixing} we
will incorporate the effects of mixing on both types of final states.

\subsection{Ditau distributions}

\begin{figure}
\begin{center}
\includegraphics[width=10.0cm]{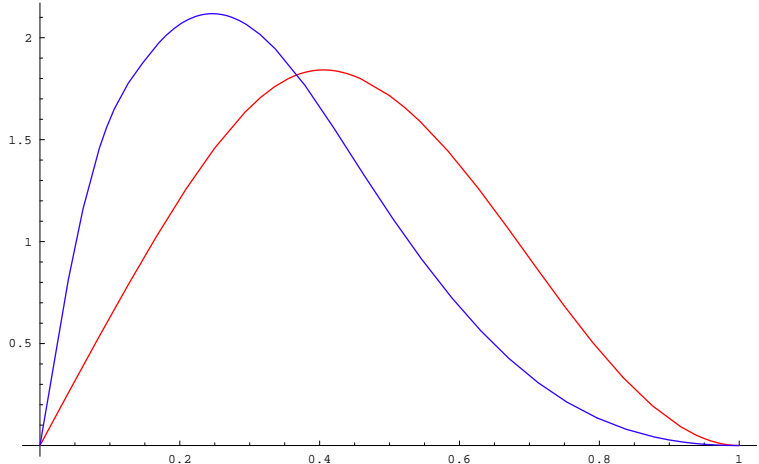}
\end{center}
\vspace{-0.75cm}
\begin{small}
\caption{Ditau triangles, for intermediate $\stau_R $ (red) and
intermediate $\stau_L $ (blue), using one-prong hadronic $\tau$s.  
The distributions are normalized to unity.  No mixing.
\label{fig:triangles}}
\end{small}
\end{figure}

Neutralino decay through staus,
\beq
\label{eq:tau-triangle}
\chii0\to\tau_{L, R} ^\pm\stau_{L, R} ^\mp\to \tau_{L, R} ^\pm\tau_{L, R} ^\mp\chij0,
\eeq
yields a triangular distribution for the ditau invariant mass in the
opposite-sign channel.  We plot the resulting observable invariant
mass distributions of the tau decay products in
figure~\ref{fig:triangles}.  These curves are a result of convolving
the underlying triangle distribution with (1) $P^-(z_1)P^-(z_2) $, for
an intermediate $\stau_L $, and (2) $P^+(z_1)P^+(z_2) $, for an
intermediate $\stau_R $.  These curves are fairly well separated and
it should be possible to clearly distinguish between these two
scenarios.

\begin{figure}
\begin{center}
\includegraphics[width=10.0cm]{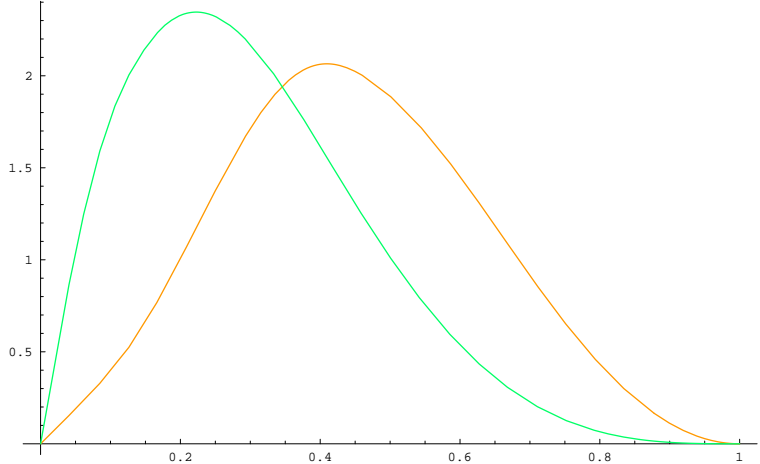}
\end{center}
\vspace{-0.75cm}
\begin{small}
\caption{The ditau hump (green) and half-cusp (orange), convolved with 
the $(+-)$ combination of energy distribution functions.  Using one-prong
hadronic $\tau$s.  No mixing.
\label{fig:hc}}
\end{small}
\end{figure}

Neutralinos intermediate between two staus lead to an underlying hump
distribution in the same-sign channel, simultaneous with an underlying
half-cusp distribution in the opposite-sign channel.  These humps and
half-cusps involve one $\tau_L$ and one $\tau_R $ and therefore the
observable distributions are obtained by convolution with the $(+-)$
combination of tau energy transfer functions.  The resulting
distributions are plotted in figure \ref{fig:hc}.  These distributions
must have equal normalization and endpoints.  While in practice the
upper endpoint may be difficult to discern, this nonetheless
translates into a stringent correlation on the relative locations of
the peaks in the opposite-sign and same-sign distributions.

\begin{figure}
\begin{center}
\includegraphics[width=10.0cm]{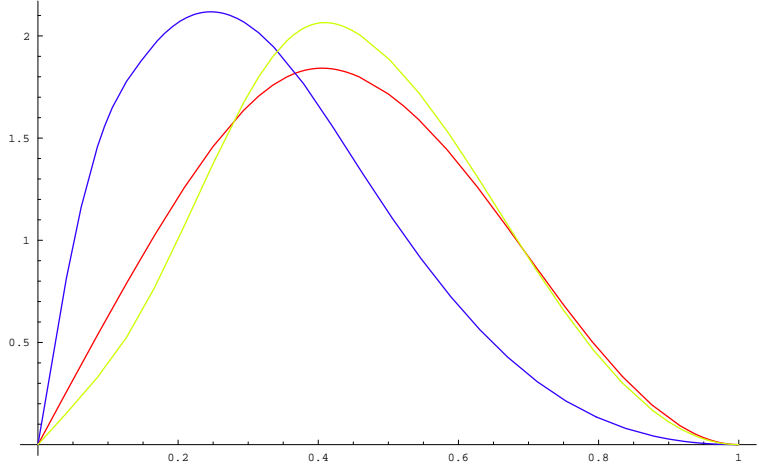}
\end{center}
\vspace{-0.75cm}
\begin{small}
\caption{Opposite sign ditau distributions: both possible triangle
distributions (with tau polarizations $(++)$ in red and $(--) $ in blue), plotted against 
the half-cusp (with tau polarizations $(+-)$, in orange). Using one-prong hadronic
$\tau$s.  No mixing.
\label{fig:os}}
\end{small}
\end{figure}

An important question is at what level the possible opposite-sign
distributions can be distinguished from each other.  Toward this end
we plot both possible triangle distributions and the half-cusp in
Figure~\ref{fig:os}.  While with enough statistics the three curves 
might possibly be distinguished, on a practical level generating and
testing hypotheses to explain an opposite-sign ditau signal will
proceed first by cross-channel comparisons.  First, if the
opposite-sign ditau distribution is an underlying half-cusp, then
there should be a corresponding underlying hump distribution in the
same-sign channel.  The absence of a same-sign signal strongly
suggests that the opposite-sign signal is an underlying triangle
distribution, due to intermediate staus.  In addition, opposite-sign
triangles and opposite-sign half-cusps fit into hypotheses which make
different predictions for object counts in the rest of the signal
events.  In particular, if the opposite-sign signal is due to
half-cusps, then there should be a larger number of leptons in the
event (most likely $\tau$'s) coming from decays into the $\stau$
initiating the decay chain, and from subsequent decay of the $\stau$
terminating the decay chain.  These additional $\tau$'s will naturally
present some combinatorial complications, which again we will not
address here.

\subsection{Lepton-tau distributions}
\label{sec:lepton-tau}

\begin{figure}
\begin{center}
\includegraphics[width=10.0cm]{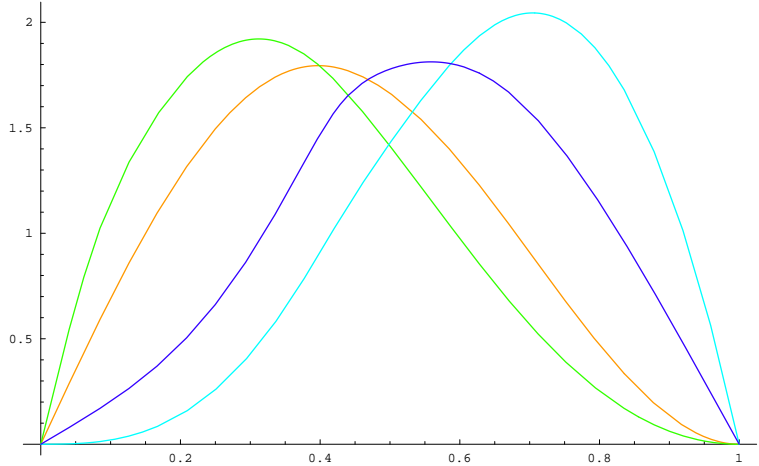}
\end{center}
\vspace{-0.75cm}
\begin{small}
\caption{All four possible lepton-tau distributions: the hump with
negative (green) and positive (orange) polarizations, the cusp with
negative (blue) and positive (cyan) polarizations. Using one-prong
hadronic $\tau$s.  No mixing.
\label{fig:lt-4}}
\end{small}
\end{figure}

Lepton-tau distributions arise from slepton decay to a stau through a
neutralino,
\[
 \slepton_{L,R} \to \ell \chii0 \to\ell\tau \stau_{L, R}, \ell\tau \stau_{R, L},
\]
or the analogous process with initial stau and final slepton.  In the
absence of mixing, there are four possible observable distributions in
lepton-tau channels, namely the hump convolved with the $(+)$ energy
distribution function; the hump convolved with the $(-)$ energy
distribution function; the half-cusp convolved with the $(+)$ energy
distribution function; and the half-cusp convolved with the $(-)$
energy distribution function.  These distributions are plotted in
Figure~\ref{fig:lt-4}.  Depending on the handedness of both the
slepton and the stau participating in the decay chain, all of these
shapes may appear in either the same-sign channel or the opposite-sign
channel, as we summarize in Table \ref{table:lt}.  Note that, for any
given process, there is a specific prediction for which of the four
shapes must appear in the same-sign channel and which must appear in
the opposite-sign channel.  They must have the same normalization, and
the same endpoints.  In addition it is worth pointing out that the
helicity of the tau in both the opposite-sign and same-sign channels
is the same.

\begin{table}
\begin{center}
\begin{tabular}{ccc}
\hline \hline
& & \\

Process &  Same-Sign & Opposite-Sign  \\
&  & \\
 \hline
 &  &\\

$\slepton_R\to\ell\tau\stau_R $ & $C_R$ & $H_R$\\
$\slepton_R\to\ell\tau\stau_L $ & $H_L $ & $C_L $\\
$\slepton_L\to\ell\tau\stau_R $ & $H_R $ & $C_R $\\
$\slepton_L\to\ell\tau\stau_L $ & $C_L $ & $H_L $\\

& \\
\hline \hline
\end{tabular}
\label{lt}
\caption{Possible lepton-tau distributions in the absence of mixing.
Here we denote by $C_R $ the half-cusp distribution convolved with the
positive polarization energy distribution function, and by $H_L $ the
hump distribution convolved with the negative polarization energy
distribution function.  Identical conclusions pertain if the roles of
the stau and the slepton are reversed.  An experimental
determination of the channel in which the half-cusp and hump
distributions appear determines the relative handedness of the slepton
and stau.
\label{table:lt}}
\end{center}
\end{table}

It should be readily possible to distinguish whether the cusp
distribution occurs in the same-sign or the opposite-sign channel, as
even after convolution the hump and half-cusp distributions are fairly
distinct.  {\em This allows one to distinguish between the scenario
where the slepton and the stau have the same handedness, and the
scenario where the slepton and the stau have opposite handedness.}  To
proceed further one would like to identify the handedness of the stau
and therefore of the slepton.  This requires comparing the $C_R $
distribution to the $C_L $ distribution, and likewise between the $H_R
$ and the $H_L $ distributions.  While this may be challenging for the
humps, the cusps are more distinct.  The discriminatory power is
enhanced by the existence of two channels which must both have the
same polarization.  With enough statistics we expect that the identity
of the stau and therefore of the slepton can be discerned.

\section{Ditau and lepton-tau distributions in the presence of mixing}
\label{sec:tau-mixing}

In many realistic SUSY models, mixing in the tau sector is
nonnegligible.  As left- and right-handed $\tau $'s have different
daughter energy spectra, the observable invariant mass distributions
will be a weighted sum of the distributions for purely left- and
purely right-handed $\tau$s, with weights determined by the mixing
parameters.  In addition, as we saw with $b$-$\l$ distributions in
section~\ref{sec:bl}, reducing the chirality of the
fermion-sfermion-neutralino vertices serves to wash out angular
correlations from intermediate fermions.  A careful fit of observed
tau distributions therefore has the potential to probe the chiral
structure and mixings of the new physics.

The Yukawa interactions between the two stau mass eigenstates, the
$i^{th} $ neutralino, and the right and left-handed taus are
\beq
{\cal L}_{int} = \stau_1 \left( \tau_R \chii0 \; y_{1,i} ^R + (\chii0) ^\dag\tau_L ^\dag \; y^L_{1,i}\right)
                  + \stau_2 \left( \tau_R \chii0 \; y_{2,i} ^R + (\chii0) ^\dag\tau_L ^\dag \; y^L_{2,i}\right)
                  +\mathrm{H.c.} .
\eeq
(In our conventions, $\tau_R$ is a left-handed anti-tau.)  The stau-tau-neutralino
Yukawa couplings are
\barray
y_{1,i} ^R & = & \sin\theta_{\stau}\lambda_\tau U_{di}^* +\cos\theta_{\stau}{\sqrt{2} g'} U_{Bi}^* \\
y^L_{1,i}  & = & \sin\theta_{\stau}\left( -{g\over\sqrt 2} U_{iW} -{g'\over \sqrt 2} U_{iB }\right) +\cos\theta_{\stau} \lambda_\tau U_{di}  \\
y^R_{2,i,} & = & \cos\theta_{\stau}\lambda_\tau U_{di}^*-\sin\theta_{\stau}{\sqrt 2 g'} U_{Bi}^* \\
y^L_{2,i}  & = & \cos\theta_{\stau}\left( -{g\over\sqrt 2} U_{iW} -{g'\over \sqrt 2} U_{iB }\right) -\sin\theta_{\stau} \lambda_\tau U_{di}  ~,
\earray
with the property that $y^L_{2, i}\to 1, y^R_{1,i} \to 1, y^L_{1,
i}\to 0, y^R_{2,i} \to 0$ as the both the stau and neutralino mixings
are turned off, that is, as $\theta_{\stau}, U_{di} \to 0$.  The
combinations of these parameters which enter into the observable
distributions are the relative probabilities of producing right- and
left-handed taus at each vertex,
\begin{eqnarray}
\cos^2 \phi_{1,i} & \equiv & \frac{|y^R_{1, i} | ^ 2}{|y^R_{1, i} | ^ 2+ | y^L_{1, i} | ^ 2}
\label{phi1} \\
\cos^2 \phi_{2,i}   & \equiv  & \frac{|y^L_{2, i} | ^ 2}{|y^L_{2, i} | ^ 2+ | y^R_{2, i} | ^ 2}\;.
\label{phi2}
\end{eqnarray}

\subsection{Ditau triangles}
\label{sec:rescaling}

Consider first the FSF process, neutralino to stau to neutralino.
This gives a ditau ``triangle''.  It is possible now for each $\tau$
to be either positively or negatively polarized, with a probability
depending on the mixings in both the neutralino and the stau sectors.

\begin{figure}
\begin{center}
\includegraphics[width=10.0cm]{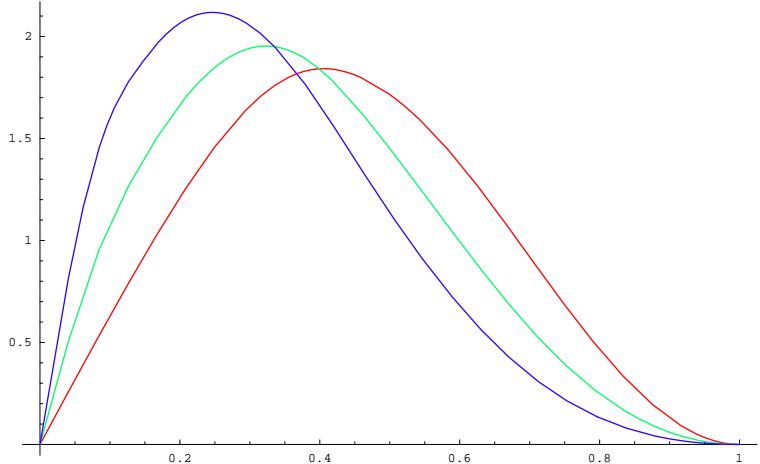}
\end{center}
\vspace{-0.75cm}
\begin{small}
\caption{All three possible ditau triangles: $T_{LL}$ (blue), $T_{LR}
= T_{RL} $ (green), and $T_{RR} $ (red). Using one-prong hadronic
$\tau$s.  All curves have the same normalization and endpoints.
\label{fig:3triangles}}
\end{small}
\end{figure}

The general distribution for this process is a sum of three possible
fundamental shapes, corresponding to the triangle distribution
convolved with the three different combinations of tau polarizations,
$T_{LL}, T_{LR}, T_{RR}$.  These shapes are plotted in
Figure~\ref{fig:3triangles}.  For general mixings, the cascade $\chij0
\to \stau_1 \tau \to \tau \tau \chii0$ now gives the invariant mass
distribution
\barray
\label{eq:triangle-three-para}
D(x) &=& \cos ^ 2\phi_{1,i}\,\cos ^ 2\phi_{1,j}\, T_{RR}(x) \\
\nonumber
     & & \:\: + (\cos ^ 2\phi_{1,i}\,\sin ^ 2\phi_{1,j}+ \sin ^ 2\phi_{1,i}\,\cos ^ 2\phi_{1,j}) T_{LR}(x) \\
\nonumber
     & & \:\: +   \sin ^ 2\phi_{1,i}\,\sin ^ 2\phi_{1,j} \,T_{LL}(x).
\earray
This defines a two-parameter family of distributions.  In principle
one could fit the observed ditau invariant mass distribution to this
formula, using a two-parameter fit.  This would measure the mixings in
the stau and neutralino sectors as well as the helicities of the taus.
Such an analysis is difficult, however, as the fit is not completely
straightforward.  First, the distribution $T_{LR}$ only differs from
the average $(T_{LL}+T_{RR})/2$ by a few percent. Second, it is
difficult in practice to accurately locate the upper endpoint of these
distributions.  We now discuss a method of analysis which deals with
these two issues.

The ditau distributions are weighted towards smaller invariant mass,
$x\lsim 0.5$, even for underlying half-cusp or triangle distributions
which peak at large invariant mass, due to the energy lost to
neutrinos.  An unfortunate consequence of this feature is that the
upper endpoints are poorly defined for many of these distributions, as
can be seen in Figures~\ref{fig:3triangles}, \ref{fig:ss-ditau}, and
\ref{fig:os-ditau}.  Anchoring the distributions by their endpoints
may then not be feasible experimentally.  The most visible feature of
all of these distributions is not the upper endpoint but rather
the location of the maximum.  Moreover, the shape and relative
location of the maxima are distinct features of the various
distributions.  A reliable way to fit experimental data to (sums of)
these distributions is thus to fit to the location of the peak,
rescaling the normalizations to preserve the total area of the
distributions---that is, the total number of events---while allowing
the relative locations of the endpoints to vary. Given a measured
ditau distribution with a peak located at invariant mass
\begin{equation}
m_{h h^{\prime}}  = m_{peak},
\end{equation}
the experimental data can be compared to theoretical distributions
rescaled to have the same peak location.

\begin{figure}
\begin{center}
\includegraphics[width=10.0cm]{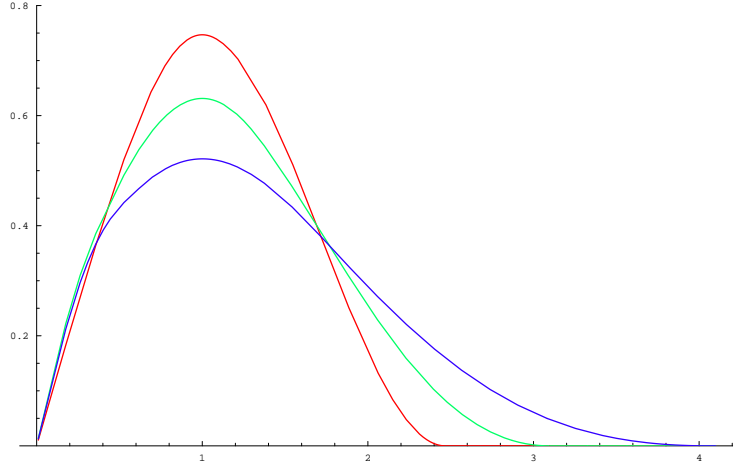}
\end{center}
\vspace{-0.75cm}
\begin{small}
\caption{The three possible ditau triangles: $T_{LL}$ (blue), $T_{LR}
= T_{RL} $ (green), and $T_{RR} $ (red), rescaled to have the same
normalization and location of peak. Using one-prong hadronic $\tau$s.
\label{fig:3-tri-scaled}}
\end{small}
\end{figure}

To do this, consider a generic ditau distribution $T_{ \{a\} } (x)$ which is a
linear combination of the basic theoretical distributions $T_{LL}$,
$T_{RR}$ and $T_{LR}$,
\[
T_{ \{a\} }(x) = a_{LL} T_{LL}(x) +a_{RR} T_{RR}(x)+a_{LR} T_{LR} (x).
\]
This distribution $T_{ \{a\} }(x)$ has a maximum at a fixed numerical value
$x_{peak}\equiv c_{ \{a\} }$, which is uniquely determined by the coefficients
$a_{LL}$, $a_{RR}$ and $a_{LR}$, Recall that the scaled variable $x$
is the ratio of the invariant mass to the endpoint,
\[
x =\frac{m_{h h^{\prime}}}{m_{end}}.
\]
If the (scaled) theoretical distribution $T_{ \{a\} } (x)$ is to describe the
data with a peak at $m_{peak}$, it must have an endpoint at $m_{end} =
m_{peak}/c_{ \{a\} }$.  Define a new variable $y$ which is scaled by the
visible location of the peak, rather than the location of the
endpoint:
\[
y \equiv \frac{x}{c_{ \{a\} }} =\frac{m_{hh'}}{c_{ \{a\} } m_{end}} =\frac{m_{hh'}}{m_{peak}}.
\]
In terms of $y$, the properly normalized theoretical distributions to
fit to data are
\begin{equation}
P_{ \{a\} }(y) = c_{ \{a\} } T_{ \{a\} }\left(c_{ \{a\} } y\right),
\end{equation}
where $\int^{1/c_{ \{a\} }}_0 dy P_{ \{a\} }(y)=1$.  By construction all distributions
$P_{ \{a\} }$ have a peak at the same location.  The three rescaled triangle
distributions are plotted in Figure~\ref{fig:3-tri-scaled}.

This rescaling procedure maximizes the distinguishability of the
different possible distributions and allows the fit to be performed
without any knowledge about the location of the endpoint in the
experimentally observed distributions.  However, it does require
backgrounds to be well-characterized, as the total number of events
needs to be well understood.

We now address the approximate degeneracy between $T_{LR}$ and
$(T_{LL}+T_{RR})/2$.  Writing $\sin^2\phi_{1, j} =w_j$,
equation~(\ref{eq:triangle-three-para}) can be rewritten
\beq
\label{eq:foo}
D(x)= T_{RR}(x)+ (w_1+w_2)(T_{LR}-T_{RR}) +w_1 w_2 (T_{RR}+T_{LL}-2 T_{LR}).
\eeq
The final term in parentheses can be neglected, subject to the limits
of experimental precision.  Moreover, this term is multiplied by a
coefficient quadratic in $w_j$.  If the angles $\phi_{1, j}$ are
small---so that the $\tau$'s being produced are predominantly
right-handed---then the final term in equation~(\ref{eq:foo}) is
doubly small.  The experimental data can then be fit to
\beq
\label{eq:footoo}
\hat D(x)= T_{RR}(x)+ (w_1+w_2)(T_{LR}-T_{RR})
\eeq
using a one-parameter fit.  This is a useful parameterisation when one
suspects that the taus being produced are mostly positively polarized,
as one would be able to learn from an examination of the relative
energy carried by charged and neutral hadrons in the reconstructed
taus.  If the taus are mostly left-handed then a more useful
parameterization of equation~(\ref{eq:triangle-three-para}) is
obtained by taking $\cos^2\phi_{1, j} = w_j$, and fitting to the
resulting approximate distribution
\beq
\check D(x)= T_{LL}(x)+ (w_1+w_2)(T_{LR}-T_{LL}).
\eeq
Of course, in an intermediate situation the parameterization
$\cos^2\phi_{1, 1} = w_1, \cos^2\phi_{1, 2}= 1- w_2$ may be more
convenient.  In all of these cases the fit measures the sum of the
relative probabilities of producing right-handed versus left-handed
taus at both vertices.

\subsection{Ditau humps and half-cusps in the presence of mixing}
\label{sec:mixingditaushumpscusps}

\begin{figure}
\begin{center}
\includegraphics[width=10.0cm]{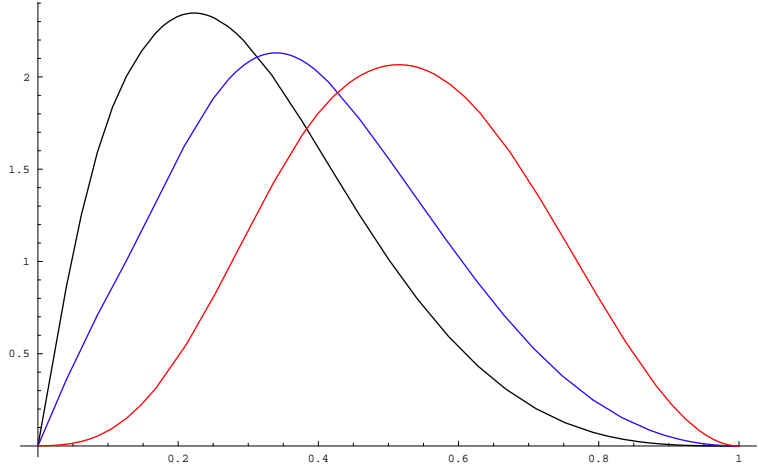}
\end{center}
\vspace{-0.75cm}
\begin{small}
\caption{The three distributions which contribute to the same-sign
ditau distribution: $H_{LR}$ (black), $C_{LL}$ (blue), and $C_{RR} $
(red). Using one-prong hadronic $\tau$s.  All curves have the same
normalization and endpoints.
\label{fig:ss-ditau}}
\end{small}
\end{figure}

\begin{figure}
\begin{center}
\includegraphics[width=10.0cm]{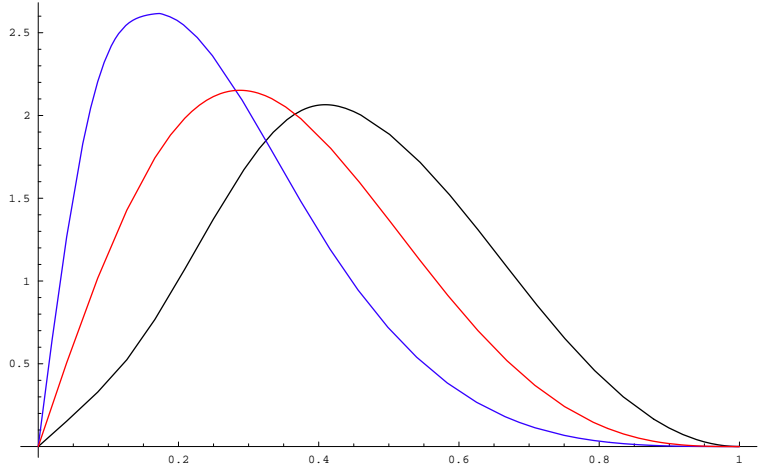}
\end{center}
\vspace{-0.75cm}
\begin{small}
\caption{The three distributions which contribute to the opposite-sign
ditau distribution: $C_{LR}$ (black), $H_{LL}$ (blue), and $H_{RR} $
(red). Using one-prong hadronic $\tau$s.  All curves have the same
normalization and endpoints.
\label{fig:os-ditau}}
\end{small}
\end{figure}

In the presence of mixing, the cascade
\beq
\label{eq:moose}
\stau_2  \to \tau\chii0\to\tau\tau\stau_1
\eeq
now gives the invariant mass distribution
\barray
\label{eq:hump-three-para}
D_{SS}(x) &=& (\cos ^ 2\phi_{1,i}\cos ^ 2\phi_{2,i}\,  +   \sin ^ 2\phi_{1,i}\sin ^ 2\phi_{2, i})H_{LR}(x) \\
\nonumber
     & & \:\: + \cos ^ 2\phi_{1,i}\sin ^ 2\phi_{2,i}\,C_{RR} + \sin ^ 2\phi_{1,i}\cos ^ 2\phi_{2,i}\, C_{LL}(x)
\earray
in the same-sign channel, and the distribution
\barray
\label{eq:cusp-three-para}
D_{OS}(x) &=& (\cos ^ 2\phi_{1,i}\cos ^ 2\phi_{2,i}\,  +   \sin ^ 2\phi_{1,i}\sin ^ 2\phi_{2, i})C_{LR}(x) \\
\nonumber
     & & \:\: + \cos ^ 2\phi_{1,i}\sin ^ 2\phi_{2,i}\,H_{RR} + \sin ^ 2\phi_{1,i}\cos ^ 2\phi_{2,i}\, H_{LL}(x)
\earray
in the opposite-sign channel.  As we have discussed above, object
counts and cross-channel correlations will likely first be used to
distinguish between FSF scenarios and SFS scenarios, so for the moment
we concentrate on the mixing-induced modifications to the ditau
distributions which result from the process (\ref{eq:moose}) alone.
The three distributions which appear in
equation~(\ref{eq:hump-three-para}) are plotted in Figure~\ref{fig:ss-ditau}.  
The distribution $H_{LR}$, which would appear in
the limit of no mixing, is shown in black, and the mixing-induced
contributions $C_{LL}$ and $C_{RR} $ are shown in blue and red
respectively.  Similarly, the three distributions which appear in
equation~(\ref{eq:cusp-three-para}) are plotted in Figure~\ref{fig:os-ditau}.  
The distribution $C_{LR}$ is shown in black, and
the mixing-induced contributions $H_{LL}$ and $H_{RR} $ in blue and
red respectively.

\begin{figure}
\begin{center}
\includegraphics[width=10.0cm]{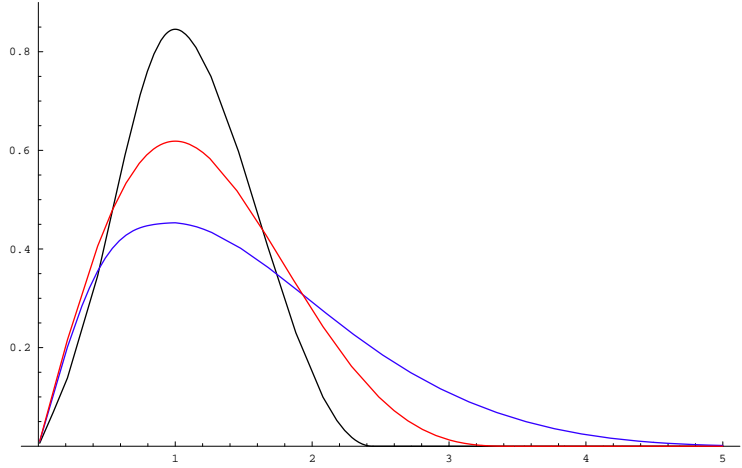}
\end{center}
\vspace{-0.75cm}
\begin{small}
\caption{The three distributions which contribute to the opposite-sign
ditau distribution: $C_{LR}$ (black), $H_{LL}$ (blue), and $H_{RR} $
(red), rescaled to have the same normalization and location of
peak. Using one-prong hadronic $\tau$s.
\label{fig:os-scaled}}
\end{small}
\end{figure}

\begin{figure}
\begin{center}
\includegraphics[width=10.0cm]{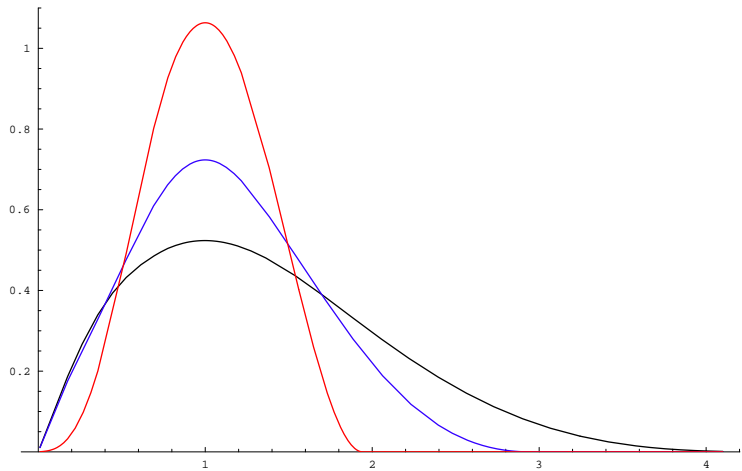}
\end{center}
\vspace{-0.75cm}
\begin{small}
\caption{The three distributions which contribute to the same-sign
ditau distribution: $H_{LR}$ (black), $C_{LL}$ (blue), and $C_{RR} $
(red), rescaled to have the same normalization and location of peak.
Using one-prong hadronic $\tau$s.
\label{fig:ss-scaled}}
\end{small}
\end{figure}

To compare the distributions to data we rescale the distributions to
fit the location of the peak and the total number of events, as with
the triangles.  The rescaled distributions are plotted in Figures
\ref{fig:os-scaled} and \ref{fig:ss-scaled}.  Unlike the triangles,
all three curves in each channel are distinct enough to allow the
possibility of a full two-parameter fit, although high
statistics would be required.  In addition one is able to perform the
same measurement in both the opposite-sign and same-sign channel.  In
this case the probabilities of producing right- and left-handed taus
at each vertex could separately be measured, in principle.  In
practice, any measurement of this sort would require a good strategy for
dealing with the combinatorial complications of additional $\tau $s in
the event, as well as the likely presence of opposite-sign ditau
triangle distributions in addition to the opposite-sign (mixed)
half-cusp of (\ref{eq:cusp-three-para}).

\subsection{Lepton-tau distributions with mixing}

The possible lepton-tau distributions in the absence of mixing are
listed in Table~\ref{table:lt}.  Once mixing is turned on, a channel
which began with a hump (half-cusp) distribution and a given
polarization in the absence of mixing will also have a contribution of
the half-cusp (hump) distribution, convolved with the opposite
polarization.  Thus, for instance, the process
\[
\slepton_R \to \l\tau\stau_1
\]
gives the distribution
\[
D_{OS} = \cos ^ 2\phi_{1, i} \,H_R (x) +\sin ^ 2\phi_{1, i}\, C_L (x)
\]
in the opposite-sign channel, and the distribution
\[
D_{SS} =\cos ^ 2\phi_{1, i}\, C_R (x) +\sin ^ 2\phi_{1, i}\, H_L (x).
\]
As there is only one $\tau $ vertex, this is only a one-parameter fit.
The result of the fit is the relative probability of emitting right-
versus left-handed taus at the tau-stau-neutralino vertex; the
opposite-sign and same-sign channels independently constrain this
probability.  Notice that the lepton-tau distributions appear in
pairs: in any given channel, the distribution must be a weighted sum
either of $H_R (x) $ and $C_L (x) $ or of $H_L (x) $ and $C_R(x) $.
There remains the discrete ambiguity of which of these pairs of
distributions occurs in the same-sign and which in the opposite-sign
channel.  However, as the lepton-tau distributions are more distinct
from one another, and as less energy is lost to neutrinos, one
scenario should be clearly preferred.  

\begin{figure}
\begin{center}
\includegraphics[width=10.0cm]{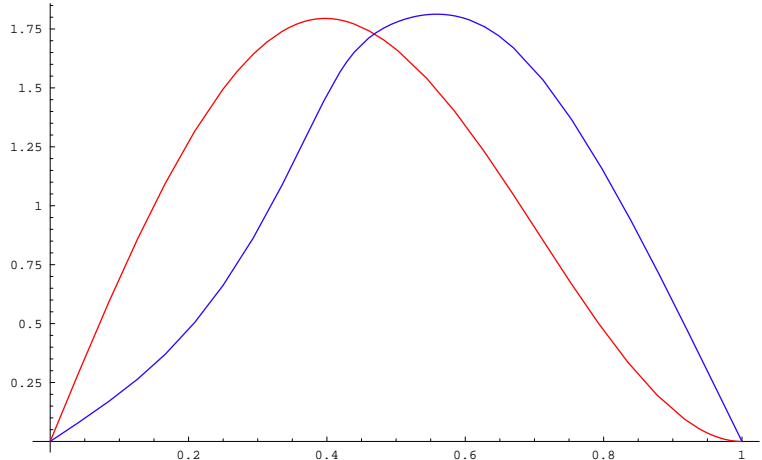}
\end{center}
\vspace{-0.75cm}
\begin{small}
\caption{The distributions $H_R (x) $ (red) and $C_L (x)$ (blue).
Using one-prong hadronic $\tau$s.
\label{fig:hpcm}}
\end{small}
\end{figure}

\begin{figure}
\begin{center}
\includegraphics[width=10.0cm]{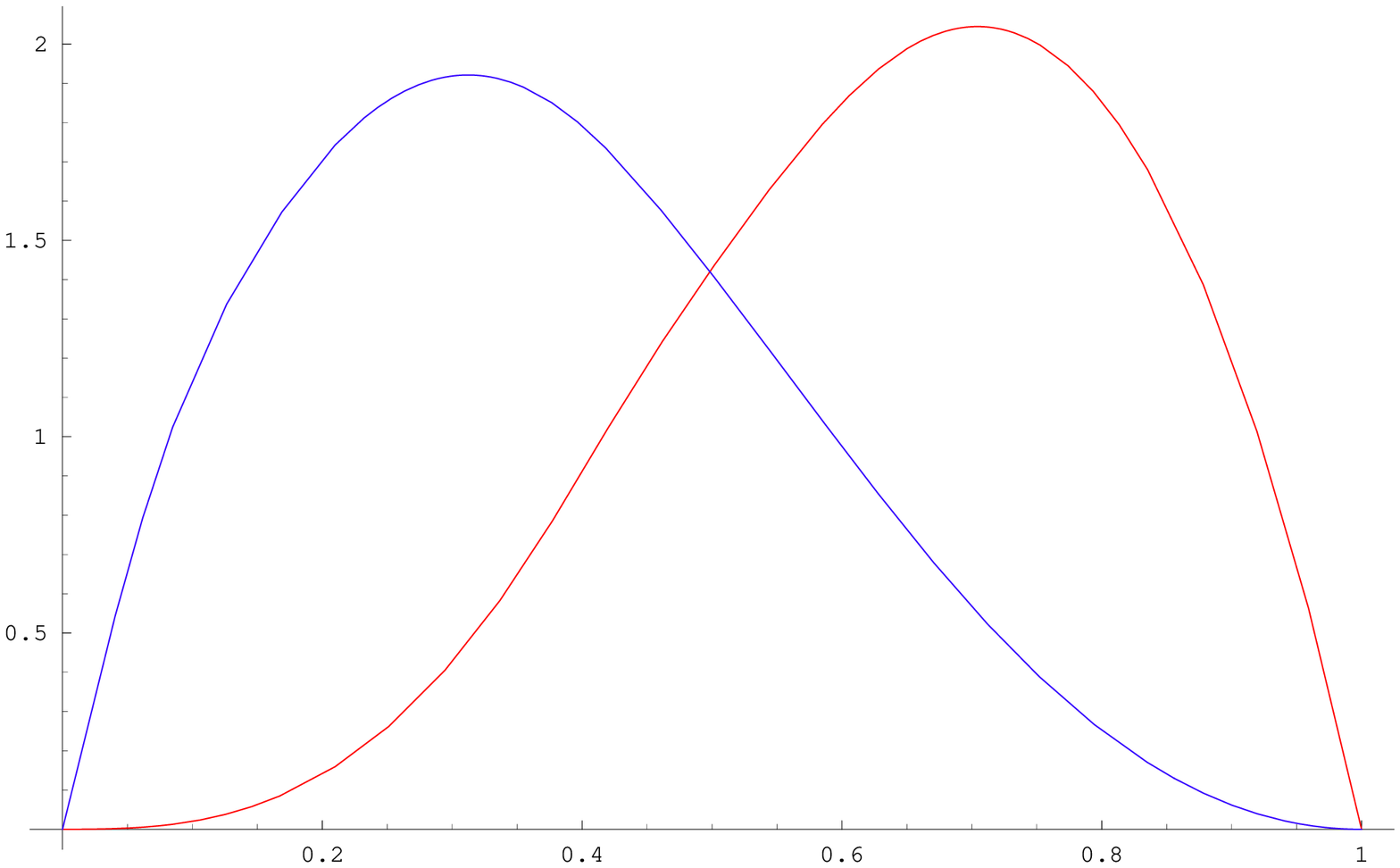}
\end{center}
\vspace{-0.75cm}
\begin{small}
\caption{The distributions $C_R (x) $ (red) and $H_L (x)$ (blue).
Using one-prong hadronic $\tau$s.
\label{fig:hmcp}}
\end{small}
\end{figure}

\begin{figure}
\begin{center}
\includegraphics[width=10.0cm]{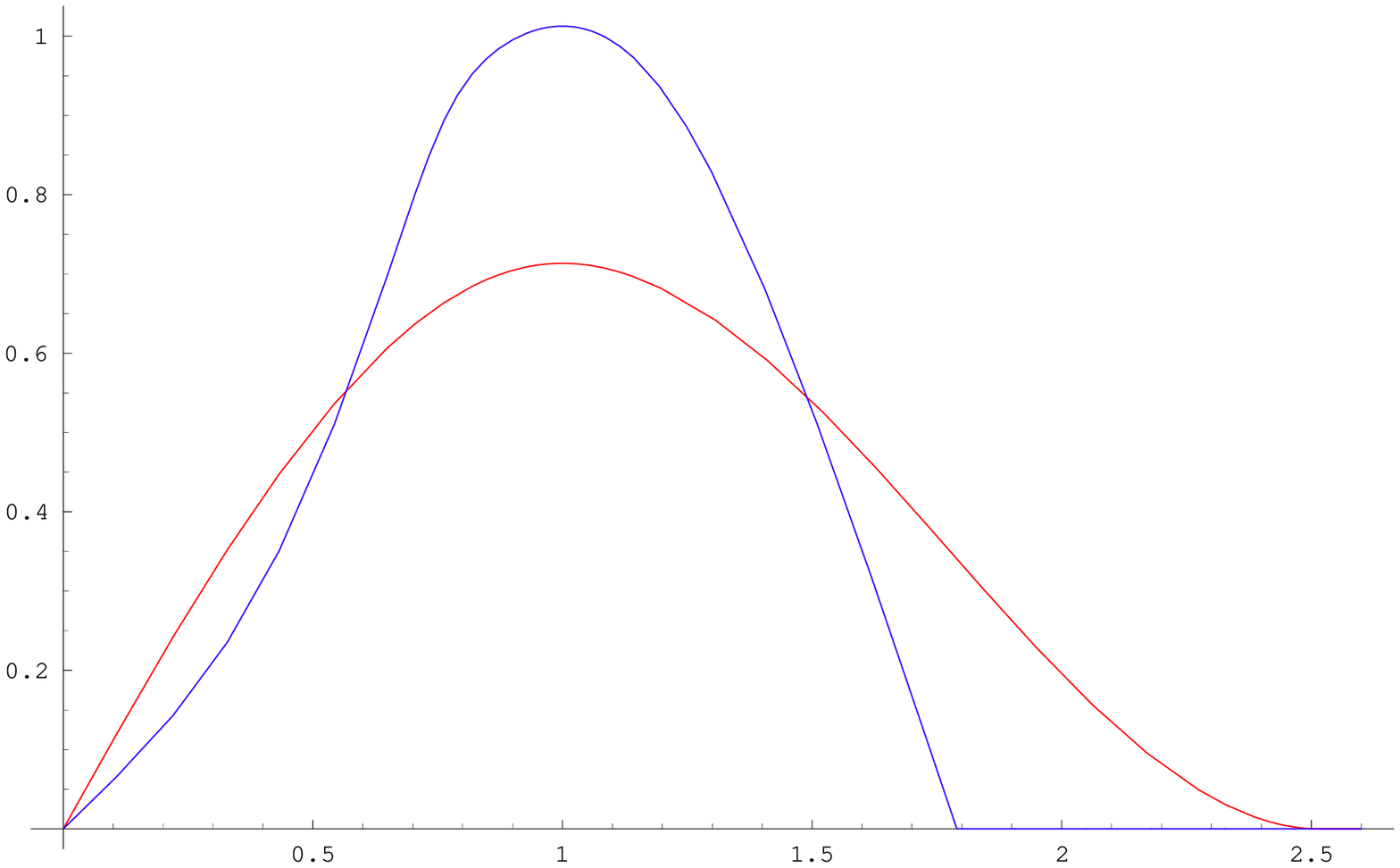}
\end{center}
\vspace{-0.75cm}
\begin{small}
\caption{The distributions $H_R (x) $ (red) and $C_L (x)$ (blue),
rescaled to have the same normalization and location of peak. Using
one-prong hadronic $\tau$s.
\label{fig:ltscaled1}}
\end{small}
\end{figure}

\begin{figure}
\begin{center}
\includegraphics[width=10.0cm]{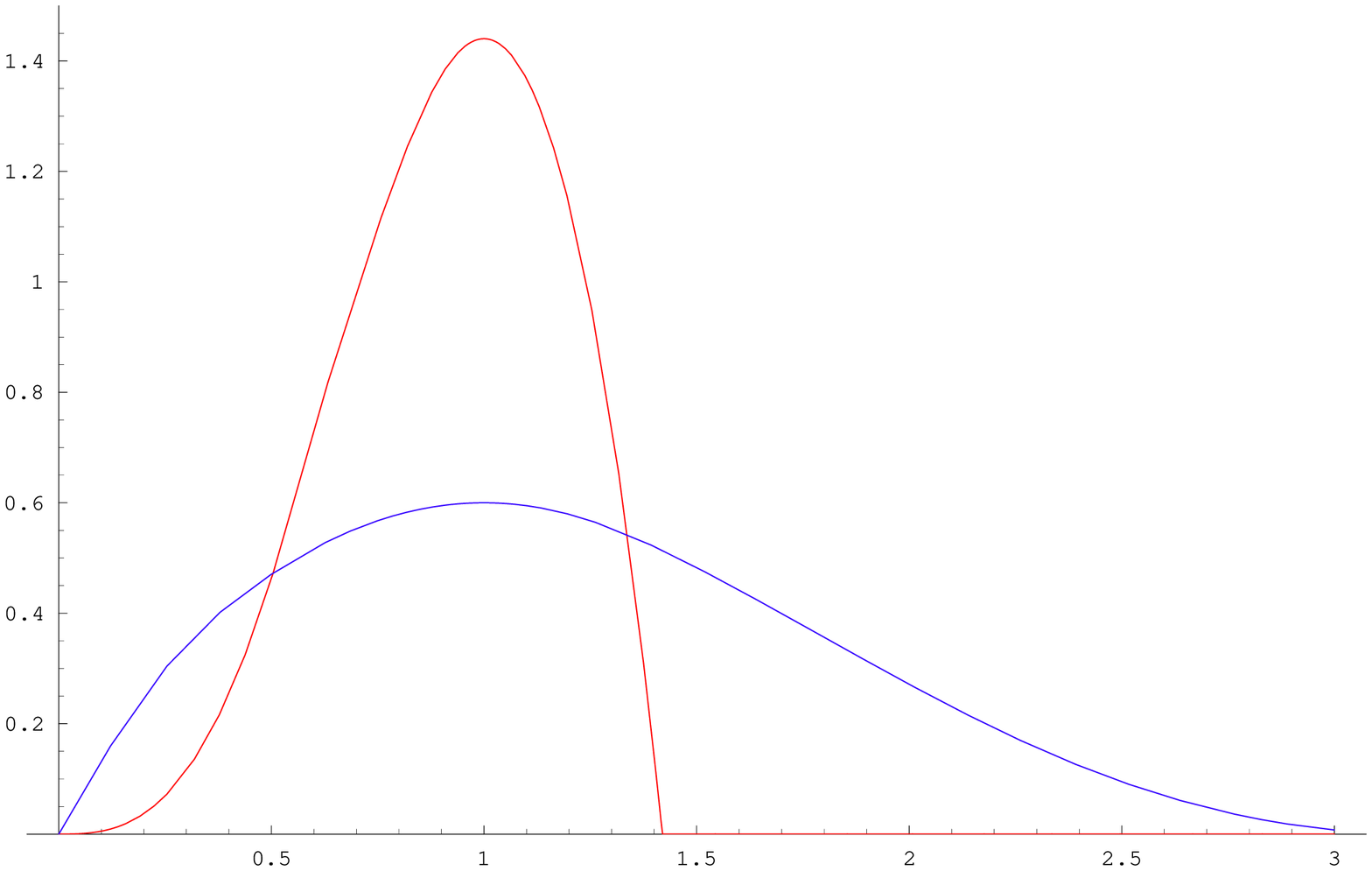}
\end{center}
\vspace{-0.75cm}
\begin{small}
\caption{The distributions $C_R (x) $ (red) and $H_L (x)$ (blue),
rescaled to have the same normalization and location of peak. Using
one-prong hadronic $\tau$s.
\label{fig:ltscaled2}}
\end{small}
\end{figure}

In Figures~\ref{fig:hpcm} and~\ref{fig:hmcp} we plot the two separate
combinations of lepton-tau distributions which can appear in any given
channel.  Compared to the ditau distributions, the peaks of these distributions
are broader, and the endpoints of these distributions are more distinct.
In these channels, fits which anchor the endpoint of the distribution rather 
than the peak may yield better results.  For comparison we plot the 
rescaled distributions with identical peak locations in
Figures~\ref{fig:ltscaled1} and~\ref{fig:ltscaled2}.

\section{Conclusions}
\label{sec:conclusion}

We have categorized the possible $b$-$\l$, $\tau $-$\l $, 
and $\tau $-$\tau $ invariant mass distributions which can arise from
two-step on-shell SUSY cascade decays and outlined several ways in 
which the special properties of $b $'s and $\tau $'s can be used to 
probe detailed properties of the underlying MSSM Lagrangian.

In the MSSM, $b$-$\l $ distributions arise from the decay of a squark through a
neutralino or chargino to a slepton.  The interplay between the nontrivial spin
of the intermediate fermion and the chiral vertices leads to a rich structure
of invariant mass distributions.  The distinctive structure of the invariant mass
distribution arising from stop decay through a chargino establishes the spin of the
chargino and depends on the mixings in the chargino-stop sector.  Signing the $b$-jet 
further allows the separate contributions of the two helicity states of a 
Majorana neutralino to be resolved, and opens the door to measurements of the 
neutralino spin and mixings in the sbottom-neutralino sector.  

Invariant mass distributions for final state $\tau$s depend on the
polarization of the $\tau$.  For analyses using invariant mass distributions 
only, hadronic one-prong $\tau$s provide the greatest sensitivity to $\tau$ 
polarization.  We have used a semi-analytic model of the one-prong decay mode
incorporating $\tau$ decays to $\pi, \rho$, and $a_1$, which together account
for $90\%$ of all one-prong $\tau$'s.  With this model we computed the possible
theoretical distributions for the observable $\tau $-$\l $ and $\tau $-$\tau $
invariant masses.

The ditau triangle distribution has excellent prospects for yielding detailed 
measurements under realistic conditions, due in part to the minimal combinatorial 
confusion in such events.  Precision fits of the ditau triangle 
can directly establish the handedness of the intermediate stau and measure
mixings in the stau-neutralino sector.  This fit
is not completely straightforward as, due to the two missing neutrinos, the
upper endpoint of the ditau triangle distribution is difficult to discern.  
The peak of the ditau triangle is more readily located, and we have proposed an 
algorithm which fits ditau distributions to the location of the peak, 
rather than the location of the endpoint.

The distributions which we 
have examined here are uniquely sensitive to spins and mixings and present several
interesting possibilities for measurements.  Much more work is necessary, however, 
in order to assess when and how well such measurements can actually be made in realistic 
circumstances.  One of the chief obstacles to fully utilizing the correlations 
identified here is the combinatorial confusion associated with identifying the 
correct pairing of objects in an event.  Without a good strategy to minimize
this ambiguity, the ability of (e.g.) $b$-$\l$ invariant mass distributions to
determine neutralino spin is greatly reduced.  As an acceptance price 
must be paid for signing the $b$-jets, cuts to reduce combinatorial 
background must be applied with care in order not to run out of signal.  However,
there are several possible approaches to reducing the combinatorial background and
we are optimistic about the possibilities of performing interesting spin and mixing 
measurements under realistic conditions.  We intend to return to this point in future work.

In our systematic consideration of $b$-$\l$, $\tau $-$\l $, and $\tau $-$\tau $ 
invariant mass distributions, we have mapped out the space of possibilities for
what shapes and intercorrelations of these distributions can appear, 
and for which combinations of parameters these distributions can probe.
We consider it instructive to understand which of these many possibilities
have been realized in benchmark models, and hope that the present work can help
guide future studies.

%
%

\section{Appendix}
\label{sec:tau}

$\tau$ decays to a single charged particle (plus any number of neutral
particles) constitute $84.7 \% $ of all $\tau$ decays.  Included in
this $84.7 \% $ is the leptonic branching fraction, $35.2 \% $ of all
$\tau$ decays, into the final states $e \bar\nu_e \nu_\tau$ and $\mu
\bar\nu_\mu \nu_\tau$.  The remaining $49.5\% $ of all $\tau$ decays
are hadronic one-prong $\tau $s.  Three-prong $\tau $'s contribute
another $15.2 \% $.  All branching fractions are taken from the PDG
\cite{pdg}.

Hadronic one-prong $\tau $s offer the most sensitive polarimeter for
polarization analyses using invariant mass distributions and are the
focus of the present work.  The dominant contributions to the hadronic
one-prong decay mode come from

\begin{itemize}
\item $\tau^- \to \nu \pi^ -$: the branching fraction into this mode
 is $11.1\% $.

\item $\tau^- \to \nu \rho ^-\to\nu \pi^- \pi^ 0$: the branching
 fraction into this mode is $25.4 \% $.

\item $\tau^- \to \nu a_1 ^-\to \nu \pi^- \pi^ 0 \pi ^ 0$: the
 branching fraction into this mode is more difficult to estimate.  By
 isospin, this mode is related to the three-prong decay mode $\tau^-
 \to \nu a_1 ^-\to \nu \pi^- \pi^ - \pi ^ +$.  The total branching
 fraction into this final state, with no additional neutrals, is
 approximately $9 \% $.  We take this to be our estimate of the total
 branching fraction of $\tau^- \to \nu a_1 ^- \to \nu \pi^- \pi^ 0 \pi
 ^ 0$.
\end{itemize}
These three decay modes contribute a branching fraction of $45\% $ to
all tau decays, leaving a total branching fraction of $5 \% $ into
one-prong taus from other modes, e.g. modes including kaons and
continuum modes with no identifiable intermediate hadronic resonance.
We model the hadronic one-prong $\tau $s as the sum of the
contributions from the three decay modes listed above.  This
incorporates $90 \% $ of the contributions to the one-prong mode, and
provides a good approximation to the full hadronic one-prong $\tau$s.
Incorporating the intermediate vector mesons $\rho, a_1$ is important
as the thresholds associated with the masses and widths of these
particles lead to notable features in the visible hadronic energy
spectra of polarized $\tau$s.

A full experimental study would be done with TAUOLA \cite{tauola},
which incorporates all measured contributions to the $\tau $ decay
modes.

The matrix elements for the process $\tau\to\nu\pi $ depend on the
polarization $\pm $ of the parent $\tau $.  They are given by (we drop
overall constants)
\begin{eqnarray}
\check\mc{M}_{+} &=&\cos\frac{\theta}{2} \\
\check\mc{M}_{-} &=&\sin\frac{\theta}{2},
\end{eqnarray}
where $\theta $ is the angle between the pion and the $\tau$ axis of
polarization in the $\tau $ rest frame.  Squaring the matrix element
and integrating over final state phase space, we obtain
\[
\frac{1}{\Gamma} \frac{d \Gamma}{d\cos\theta} = \frac{1}{2} \left( 1+
\mc{P}_\tau \cos\theta\right),
\]
where $\mc{P}_\tau = \pm$ is the polarization of the $\tau$.  These
angular distributions can be simply understood by appealing to angular
momentum conservation: the left-handed neutrino must carry off the
total angular momentum of the $\tau$, and is therefore emitted forward
for initial left-handed $\tau$s, and backward for initial right-handed
$\tau $s.

The $\tau $s produced at colliders are typically highly boosted, and
the variable of interest is not $\theta $ but rather $z$, the fraction
of the $\tau$ (lab frame) energy which is carried by the hadronic
daughters,
\[
z\equiv \frac{E_d}{E_\tau}.
\]
In the collinear approximation this quantity is invariant under
boosts along the $\tau$'s direction of motion.  In terms of the
(unknown) boost $\beta$ between the $\tau $ rest frame and the lab
frame, 
\beq
\label{eq:costoz}
\cos\theta =\frac{2 z-1-(m ^ 2_\pi/m ^ 2_\tau)}{\beta (1-(m ^ 2_\pi/m ^
2_\tau))}.
\eeq
We will henceforth work in the collinear limit, $\beta\to 1$.  Taking
this limit and dropping the factor $ m ^ 2_\pi/m ^ 2_\tau$, we obtain
for the pion spectra
\[
P ^{\pm}_{(\pi)} (z) = 1 \pm (2 z-1) .
\]

The decay modes $\tau^- \to \nu \rho ^-$, $\tau^- \to \nu a_1 ^-$ are
more complicated, due to the different contributions from the
longitudinal and transverse polarizations of the vector mesons, and
the finite widths of the intermediate states.

The matrix elements governing $\tau$ decay to $\nu v^\mu$ depend on
the polarization $\pm $ of the parent $\tau$ as well as the
polarization $T, L$ of the vector meson $v ^\mu $.  These matrix
elements are (we again drop overall constants independent of $m_v$)
\begin{eqnarray}
\label{eq:mv}
\check\mc{M}_{-T} =\sqrt{2 (m ^ 2_\tau-m ^ 2_v)}\cos\frac{\theta}{2}
\phantom{sss} &
   \check\mc{M}_{-L} =\frac{m_\tau}{m_v}\sqrt{(m ^ 2_\tau-m ^
2_v)}\sin\frac{\theta}{2} \\
\nonumber
\check\mc{M}_{+T} =-\sqrt{2 (m ^ 2_\tau-m ^ 2_v)}\sin\frac{\theta}{2}
\phantom{sss} &
    \check\mc{M}_{+L} =\frac{m_\tau}{m_v}\sqrt{(m ^ 2_\tau-m ^
         2_v)}\cos\frac{\theta}{2}
\end{eqnarray}
where $\theta $ is the angle between the vector meson and the $\tau$
axis of polarization in the $\tau $ rest frame.  Note the longitudinal
polarization $L$, which carries zero angular momentum along the tau
polarization axis in the tau rest frame, contributes like the scalar
pion, while the transverse polarization $T $ contributes oppositely.
(Note also that angular momentum conservation allows only one of the
two transverse polarizations to contribute to the decay.)  In the
narrow width approximation, the net distribution of events in $\cos
\theta$ is then of the form \cite{Hagiwara:1989fn}
\[
\frac{1}{\Gamma}\frac{d \Gamma}{d\cos\theta} =\frac{1}{2}\left(
      1+\mc{P}_\tau \left(\frac{ m ^ 2_\tau-2 m ^ 2_v}{m ^ 2_\tau+2 m
      ^ 2_v}\right)\cos\theta\right).
\]
In other words, the two different vector meson polarizations make
contributions to the $\tau$ helicity-dependent portion of the
amplitude which tend to cancel.  The factor $(m ^ 2_\tau-2 m ^ 2_v)/(m
^ 2_\tau+2 m ^ 2_v)$ is approximately $0.45$ for the $\rho $, and only
$0.03$ for the $a_1$.  Thus, with no discrimination between the
separate contributions of longitudinal and transverse $a_1$s, the
$a_1$s do not yield much sensitivity to $\tau$ polarization.

In order to obtain a realistic spectrum of hadronic daughter energies,
the finite widths of the intermediate vector mesons $\rho $, $a_1$
must be taken into account.  Following \cite{Bullock:1992yt}, we use a
Breit-Wigner distribution for the vector mesons, incorporating a
running width $\Gamma_v(m ^ 2) $, as we now detail.

In general the contribution to the decay distribution from the
intermediate vector meson $v $ takes the form
\[
 d\Gamma\propto\int d\Pi_2 (\tau\to\nu v) \; dm^2_v\; d\Pi_n (v\to n\pi) \;
    |\check\mc{M}_\mu (\tau\to\nu v) \mc{P}^{\mu\nu}_v(m ^
 2_v)\hat\mc{M}_\nu(v\to n\pi)|^2.
\]
Here $\mc{P}^{\mu\nu}_v(m ^ 2_v)$ is the vector meson propagator.  It
is convenient to write the total matrix element for this process as a
sum of contributions from the longitudinal and transverse
polarizations of the intermediate vector meson $v$.  After integrating
over the $n $-pion phase space, the interference between the vector
meson polarizations vanishes and the decay rate given above greatly
simplifies.  It is natural to define the lineshape (or vector meson
decay rate)
\[
 g_v (m ^ 2_v) = \int d\Pi_n (v\to n\pi) \; |\hat\mc{M}(v\to n\pi)|^2,
\]
which does not depend on polarization.  By the optical theorem the
lineshape is proportional to the running width of the vector meson,
\[
 m_v\Gamma_v (m ^ 2_v)\propto g_v (m ^ 2_v).
\]
The decay distribution can now be written
\[
 d\Gamma\propto\int d\Pi_2 (\tau\to\nu v) \; dm^2_v\; \left(|\check\mc{M}_L
      (\tau\to\nu v_L)| ^ 2+
                |\check\mc{M}_{T} (\tau\to\nu v_T)| ^ 2\right) D_v (m ^ 2_v)
      g_v (m ^ 2_v),
\]
where $D_v (m ^ 2_v)$ is the Breit-Wigner with a running width,
\barray
 D_v (m ^ 2_v) &= & \left[(m ^ 2_v-m ^ 2_0) ^ 2 + (m_v \Gamma_v) ^
 2\right]^{-1}\\
              &= & \left[(m ^ 2_v-m ^ 2_0) ^ 2 + (m_0 \Gamma_0)^2 \times
 \left(\frac{g_v (m_v ^ 2)}{g_v (m_0 ^ 2)}\right)^ 2 \right]^{-1}.
\earray
For the $\rho $, we take $m_0$ and $\Gamma_0$ to be
\[
  m_{0,\rho} = 770 \:\mathrm{MeV}, \phantom{spacer} \Gamma_{0,\rho} = 150 \:\mathrm{MeV}.
\]
For the $a_1 $, we take  $m_0$ and $\Gamma_0$ to be
\[
  m_{0,a} = 1.22 \:\mathrm{GeV}, \phantom{spacer} \Gamma_{0,a} = 420 \:\mathrm{MeV}.
\]
For the decay $\rho \to 2\pi$, the simple chiral Lagrangian
interaction
\[
 \mc{L}_{int} \propto \rho^\mu (\pi_1 \partial_\mu \pi_2 - \pi_2 \partial_\mu
 \pi_1)
\]
suffices for our purposes and gives the line shape
\beq
\label{eq:gr}
g_\rho (m ^ 2) = \mathrm{const.} \times \frac{(m ^ 2-4 m_\pi ^ 2) ^ {3/2}}{m}.
\eeq
The parameterization of $a_1$ decay, which proceeds dominantly through
$a_1\to\rho\pi\to 3\pi$, is more involved due to the multiple possible
parameterizations of intermediate hadronic resonances in the chiral
Lagrangian describing $a_1$ dynamics.  Again following
\cite{Bullock:1992yt}, we use a parameterization of the running width
$g_a (m ^ 2)$ due to \cite{Kuhn:1990ad},
\beq
 \label{eq:ga}
 g_a (m ^ 2) =\left\{ \begin{array}{ll}
                        4.1 (m ^ 2-9 m ^ 2_\pi) ^ 3 \left(1-3.3 (m ^ 2-9
 m ^ 2_\pi) +5.8 (m ^ 2-9 m ^ 2_\pi)^ 2\right) &
                                 m ^ 2 <(m_\rho+m_\pi) ^ 2\\
                        m ^ 2\left(1.623+\frac{10.38}{m ^
 2}-\frac{9.32}{m ^ 4} +\frac{0 .65}{m ^ 6}\right) &
                                 m ^ 2\geq (m_\rho+m_\pi) ^ 2
                       \end{array}\right.
\eeq
All masses are in units of GeV.  This particular parameterization
follows from a model of $a_1$ decay which goes through intermediate
finite-width $\rho$ alone, with no contribution from other resonances
such as the radial excitations $\rho ', \rho''$.  The coefficients are
obtained from a fit to experimental data \cite{Kuhn:1990ad}.

\begin{figure}
\begin{center}
\includegraphics[width=10.0cm]{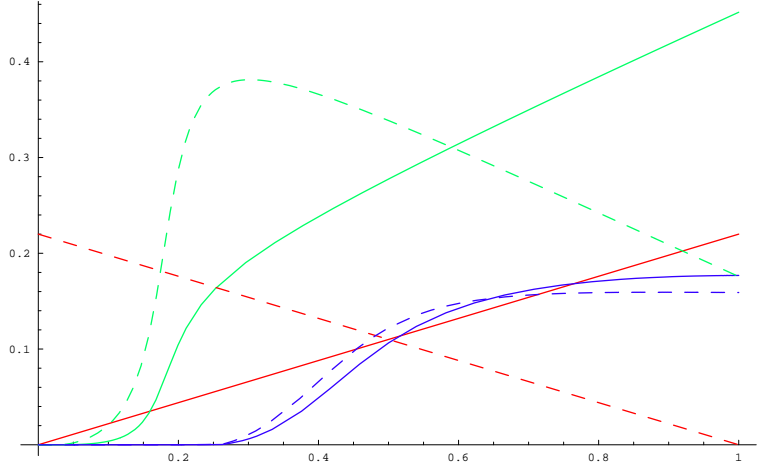}
\end{center}
\vspace{-0.75cm}
\begin{small}
\caption{Contributions to the $\tau $ daughter energy fraction
distribution for hadronic one-prong taus, weighted by branching
fraction.  The three contributing decay modes are $\tau \to \pi \nu $
(shown in red); $\tau \to \rho \nu $ (shown in green) and $\tau \to
a_1 \nu $ (shown in blue).  Solid lines indicate the distributions for
positive helicity tau, and dashed lines the contribution for negative
helicity tau.
\label{fig:breakdown}}
\end{small}
\end{figure}

Plugging in the matrix elements \er{mv} for the initial decay
$\tau\to\nu v$ and doing some simplification, we have
\[
 d\Gamma\propto\int d\cos\theta \, dm ^ 2_v\;  (m_\tau ^ 2-m_v ^ 2)^
 2\left[2+\frac{m_\tau ^ 2}{m_v ^ 2} +
          \mc{P}_\tau\cos\theta\left(\frac{m ^ 2_\tau}{m ^
 2_v}-2\right)\right]
    D_v (m ^ 2_v)\, g_v(m ^ 2_v).
\]
To change variables from $\cos\theta$ to $z $, use \er{costoz}, with $m_v$ in
place of $m_\pi $.  With this substitution, we have
\begin{eqnarray*}
 d \Gamma &\propto &\int_0 ^ 1 dz \int_{(nm_\pi) ^ 2 }^{zm_\tau ^ 2} dm_v ^ 2
 \;
                 \left[ \left(1-\frac{m_v ^ 2}{m_\tau ^ 2}\right)
 \left(2+\frac{m_\tau ^ 2}{m_v ^ 2}\right) +
          \mc{P}_\tau\left( (2 z-1)-\frac{m_v ^ 2}{m_\tau ^
 2}\right)\left(\frac{m ^ 2_\tau}{m ^ 2_v}-2\right)\right] \\
          &&\phantom{spacer spacer}\times
                D_v (m ^ 2_v)\, g_v(m ^ 2_v) .
\end{eqnarray*}
Finally then, we can write the contribution from the vector mesons as
\barray
\nonumber
\frac{1}{\Gamma} \frac{d \Gamma}{dz} &=&\mathrm{const.}\times \int_{(nm_\pi)
 ^ 2 }^{zm_\tau ^ 2} dm_v ^ 2 \;
               \left[ \left(1-\frac{m_v ^ 2}{m_\tau ^ 2}\right)
 \left(2+\frac{m_\tau ^ 2}{m_v ^ 2}\right) +
          \mc{P}_\tau\left( (2 z-1)-\frac{m_v ^ 2}{m_\tau ^
 2}\right)\left(\frac{m ^ 2_\tau}{m ^ 2_v}-2\right)\right] \\
          &&\phantom{spacer spacer}\times D_v (m ^ 2_v)\, g_v (m ^ 2_v).
\earray
Using the lineshapes \er{gr}, \er{ga} and carrying out the integral
over $m ^ 2_v$ numerically gives the hadronic energy spectra
$P^\pm_{(\rho)} (z) $, $P^\pm_{(a)} (z) $ which are plotted in
Figure~\ref{fig:breakdown}.

\bigskip

\bigskip

{ \Large \bf Acknowledgments}

\smallskip \smallskip

We would like to thank Scott Thomas for collaboration and for many
insightful conversations, as well as comments on the manuscript.  We
would also like to thank Amit Lath, Yifan Lin, Aneesh Manohar, Michael
Peskin, and Steve Schnetzer for useful discussions.  J.S. thanks the
Kavli Institute for Theoretical Physics for hospitality.  The work of
M.G. was supported by Los Alamos National Laboratory.  The work of
J.S. was supported by DOE grant DE-FG02-96ER40959.



\end{document}